\documentclass[pre,aps,twocolumn]{revtex4}
\usepackage{amsfonts}
\usepackage{amssymb}
\usepackage{graphicx}
\usepackage{mathrsfs}
\usepackage{array}
\usepackage{amsmath}
\usepackage{lscape}
\usepackage{ bbold }
\usepackage{ucs}

\begin{document}

\title{Cross-talk and transitions between multiple spatial maps in an attractor neural network model of the hippocampus: phase diagram (I)}
\author{R. Monasson, S. Rosay}
\affiliation{Laboratoire de Physique Th\'eorique de l'ENS, CNRS \& UPMC, 24 rue Lhomond,
75005 Paris, France}
\date{\today}

\begin{abstract}
We study the stable phases of an attractor neural network model, with binary units, for hippocampal place cells encoding 1D or 2D spatial maps or environments. Using statistical mechanics tools we show that, below critical values for the noise in the neural response and for the number of environments, the network activity is spatially localized in one environment. We calculate the number of stored environments. For high noise and loads the network activity extends over space, either uniformly or with spatial heterogeneities due to the cross-talk between the maps, and memory of environments is lost. Analytical predictions are corroborated by numerical simulations. 
\end{abstract}
 
\maketitle

\section{Introduction}

Understanding the representation of space by the brain is a long-lasting question, which has been adressed using many varied methods. This includes memory of places, localization of one's position, mental exploration and planning of forecoming trajectories. During the last decades, the use of microelectrodes allowing single cell recordings has revolutionized our knowledge of neural networks. In 1971, O'Keefe \& Dostrovsky \cite{OKeefeDostrovsky71} recorded neural activity in the hippocampus of rats and discovered the existence of place cells, which fire only when the animal is located in a certain position in space (called place field). This discovery suggested that hippocampus could be the support for space representation or a 'cognitive map'. Since then, many experimental and theoretical studies have been carried on hippocampus, making it one of the most, if not the most studied part of the brain \cite{HippocampusBook}.

The properties of place cells, their conditions of formation and the sensory and behavioral correlates of place fields have been investigated experimentally \cite{ThompsonBest89,Bostock91,Wilson93}. Place fields have the striking property to appear as randomly distributed, independently of the neurons locations in the neural tissue: two neighbouring neurons can have very distant place fields. Furthermore, several 'environments' or 'maps' can be learned, and a given neuron can have place fields in several environments, which are apparently randomly assigned, a property called remapping \cite{KubieMuller91}. Place fields are controlled primarily by visual cues \cite{Gothard96} but the activity of place cells persists in the dark \cite{Quirk90} and is also driven by self-motion signals, that is, 'path integration' \cite{McNaughton96}. More recently, the discovery of grid cells \cite{Fyhn04,Hafting05} in the enthorinal cortex (that feeds input into the hippocampus) opened a new way in the comprehension of a complex system of interacting brain regions \cite{Moser08}. Many theoretical models have been proposed to account for these experimental results. Beyond the comprehension of the hippocampus itself, the motivation is to reach more insights about the functional principles of the brain \cite{HippocampusBook}.

Experiments show that the hippocampus is able to learn, memorize and retrieve spatial maps. The massive intrinsic connectivity in hippocampus CA3 led to the hypothesis of an attractor neural network \cite{Rolls89,Tsodyks99} where memorized activity patterns are the attractors of the dynamics, such as in the celebrated Hopfield model  \cite{Hopfield82}. In the Hopfield model it is assumed that the patterns are additively stored in the synapses, through a Hebbian learning mechanism. A deep and quantitative understanding of the Hopfield model was made possible by the use of the statistical physics theory of mean-field spin glasses  \cite{Amit85,Amit89}. In the case of the rodent hippocampus, the memorized items are space manifolds called environments \cite{KubieMuller91}. Neural network models for place cells have been proposed allowing to perform formal computations, in particular by Battaglia \& Treves, who carried out a mean-field calculation of the storage performance of a network with linear threshold units \cite{BattagliaTreves98}. Recently Hopfield proposed a similar model for mental exploration in a network with adaptation \cite{Hopfield10}. However, the cross-talk between the different environments encoded in the network, and the transitions that can occur between them as observed experimentally \cite{Jezek11} remain poorly understood.

Here, we propose a mean-field model of interacting binary units and study the different regimes of activity in the presence of neural noise. The model is defined in Section \ref{model}. We study the case where multiple environments are memorized in Section \ref{multenvt}, and derive the different regimes of activity of the network under given conditions of neural noise and memory load in Section \ref{sectrois}. The phase diagram of the system is computed in Section \ref{phase_diagram} and compared to numerical simulations. We show that an activity of the network that is locally spatialized in one of the stored maps, as observed experimentally, is the stable state of the network provided that both the neural noise and the memory load are small enough. For high noise and/or loads the the activity is delocalized in all environments, either uniformly over space or with spatial heterogeneities controled by the cross-talk between environments (glassy phase). We finally discuss the value of the parameters (Section \ref{parameters}) and the hypothesis of the model (Section \ref{discussion}) compared to previous works. The study of the landscape and of the dynamics of the model will be addressed in a companion publication \cite{Monasson13}.

\section{The model}\label{model}

\subsection{Definition}

The $N$ place cells are modeled by interacting binary units $\sigma_i$ equal to 0 or 1, equivalent to Ising spins and corresponding to, respectively, silent and active states. We suppose that, after learning of the environment and random allocation of place fields, each place cell preferentially fires when the animal is located in an environment-specific location in the $D$-dimensional space, defining its place field. For simplicity space is assumed to be a segment of length $N$ for $D=1$, and a square of edge length $\sqrt N$ in $D=2$, with periodic boundary conditions. The $N$ centers of the place fields are assumed to be perfectly located on a $D$-dimensional regular grid: two contiguous centers are at unit distance from each other. This assumption is not necessary (heterogeneous locations of place fields in space can be considered), but allows us to remove one source of randomness and to concentrate on the interference between the stored spatial maps as one of the main sources of noise.

Let $d_c$ be the extension of a place field,  that is, the maximal distance between locations in space recognized by the same place cell. Place cells whose place fields overlap, and, therefore, spike simultaneously as the animal wanders in the environment, are assumed to strengthen their synaptic connections. Calling $d_{ij}$ the distance between the place field centers of cells $i,j$ in the environment we assume that the reinforcement process ends up with synaptic couplings given by 
\begin{equation}\label{rule1}
J^0_{ij} = \left\{\begin{array} {c c c}
\frac 1N & \hbox{\rm if} & d_{ij} \le d_c\ ,\\
0 & \hbox{\rm if} & d_{ij} > d_c \end{array} \right. \ .
\end{equation}
The positive sign of the couplings ensures that they are excitatory (ferromagnetic in spin language). We choose the place extension $d_c$ such that each cell $i$ is connected to the same number of other cells $j$, independently of the space dimension $D$. Let $w\,N$ be this number: $w(\ll 1)$ is the fraction of the neural population any neuron is coupled to. This scaling is a consequence of the assumption of place fields covering a fixed fraction of space; consequently our model is mean-field. Hence, $d_c=\frac w2 N$ in dimension $D=1$, and $d_c=\sqrt \frac{w\, N}{\pi}$ in dimension $D=2$. The $\frac 1N$ scale factor is such that the total contribution to the local field received by a place cell is finite when the number of cells, $N$, is sent to infinity. Note that we assume here that the environment is perfectly explored: couplings depend on the distance $d_{ij}$ only, and not on the particular sequence of positions occupied by the animal during the time spent in the environment. The case of partial, non-homogeneous explorations was studied in \cite{Tsodyks99}. Couplings defined by prescription (\ref{rule1})  are symmetric, and only reflect the local structure of the environment, irrespectively of the exploration process.

\begin{figure}
 \centering
   \includegraphics[width=0.45\linewidth]{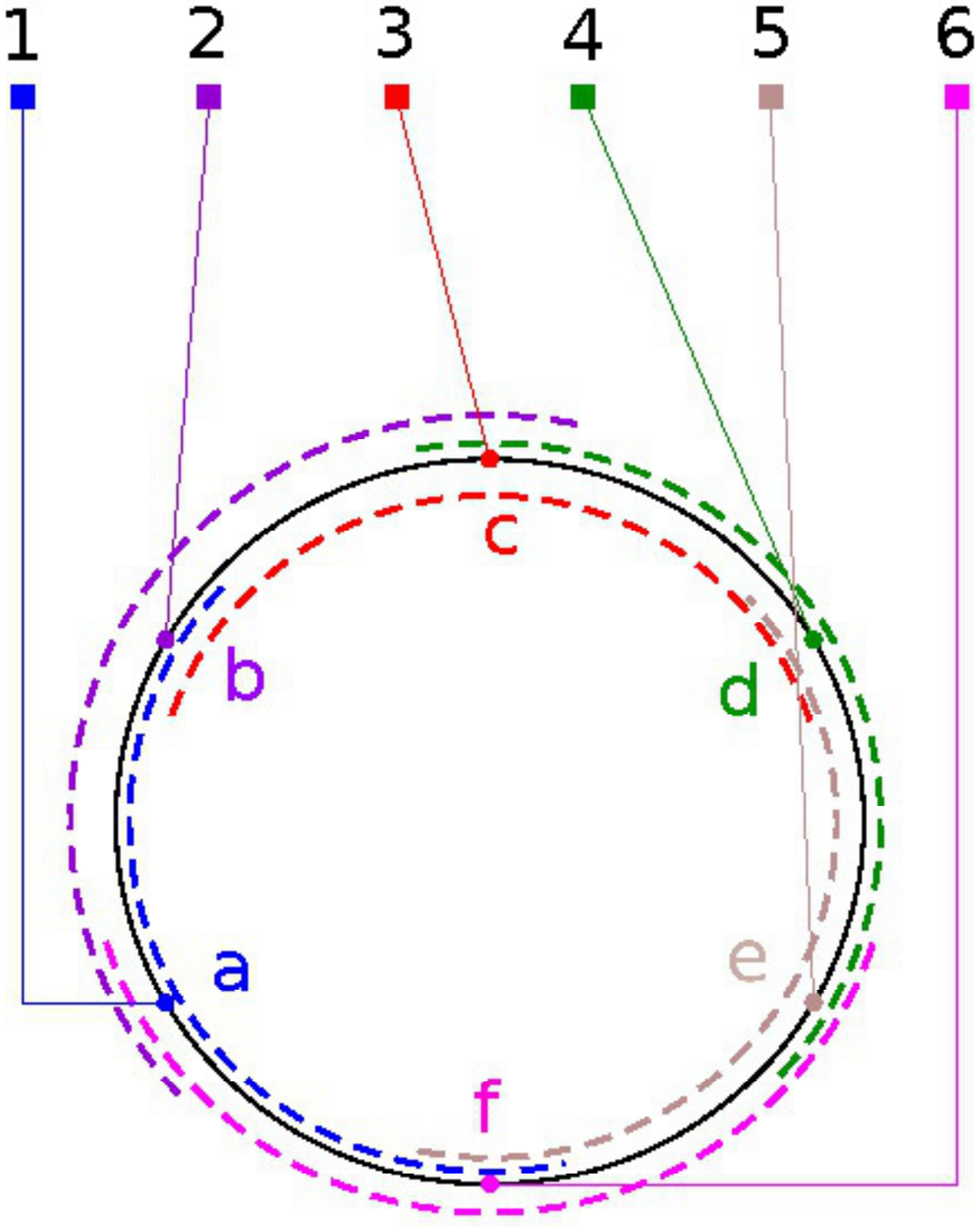}\hspace{0.05\linewidth}
   \includegraphics[width=0.45\linewidth]{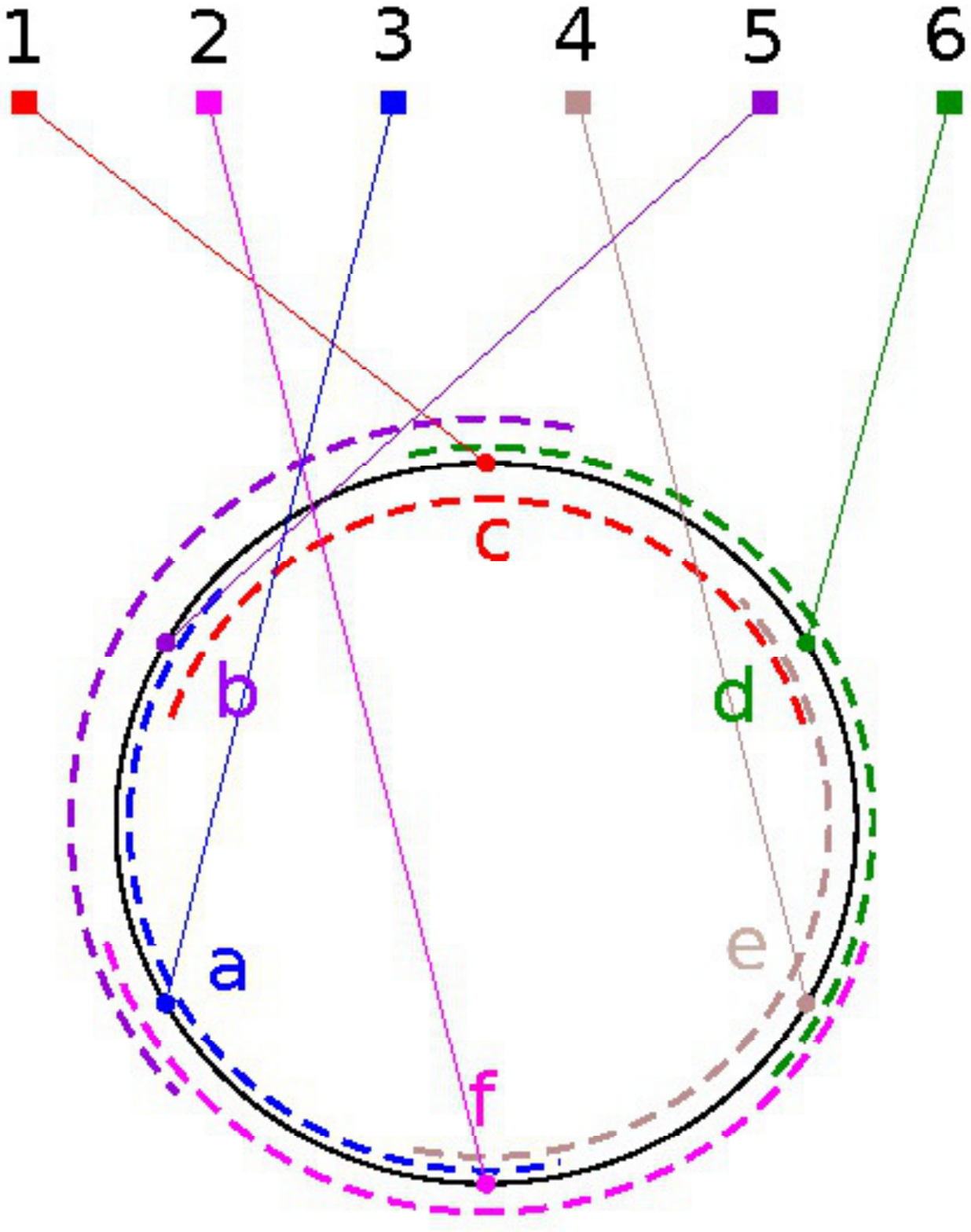}
 \caption{Example of remapping  of the place field centers of $N=6$ neurons (denoted by indices 1,..,6) in two different 1D environments with periodic boundary conditions and $w=\frac26$. Place fields in each environment are represented by colored dashed lines, place field centers are denoted by letters a,..,f. }
  \label{fig:remapping}
 \end{figure}

Each time the rodent explores a new environment a remapping of the place fields takes place. Let $L$ be the number of explored environments, in addition to the environment above (hereafter called reference environment). We assume that the remapping can be represented by a random permutation of the $N$ place-cell indices associated to the place fields in the reference environment, denoted by $\ell=0$. Let $\pi^\ell$ be the permutation corresponding to remapping number $\ell$, where $\ell=1,\dots, L$ is the index of the environment. In environment $\ell$ cells interact if the distance $d_{\pi^\ell(i)\pi^\ell(j)}$ is smaller than $d_c$, and do not interact at larger distances. An obvious modification of (\ref{rule1}) defines the coupling matrix $J^\ell$ corresponding to environment $\ell$. We finally assume that all environments contribute equally and additively to the total synaptic matrix,
\begin{equation} \label{rule2}
J_{ij}=\sum\limits_{\ell=0}^L J_{ij}^\ell = J_{ij}^0+\sum\limits_{\ell=1}^L J_{\pi^\ell(i)\pi^\ell(j)}^0 \ .
\end{equation} 
For the sake of a better understanding, we can consider an example of a matrix $J$ in the very simple case $N=6$, $w=\frac26$, $L+1=2$ and $D=1$, illustrated in figure  (\ref{fig:remapping}). For the reference environment the coupling matrix is
\begin{equation}
J^0=\frac 16 \left( 
\begin{array}{c c c c c c}
0 & 1 & 0 & 0 & 0 & 1 \\
1 & 0 & 1 & 0 & 0 & 0 \\
0 & 1 & 0 & 1 & 0 & 0 \\
0 & 0 & 1 & 0 & 1 & 0 \\
0 & 0 & 0 & 1 & 0 & 1 \\
1 & 0 & 0 & 0 & 1 & 0 
\end{array}\right)
\end{equation}

For another environment obtained through the random permutation $\pi=(3,6,1,5,2,4)$ we obtained the coupling matrix

\begin{equation}
J^1=\frac 16 \left( 
\begin{array}{c c c c c c}
0 & 0 & 0 & 0 & 1 & 1 \\
0 & 0 & 1 & 1 & 0 & 0 \\
0 & 1 & 0 & 0 & 1 & 0 \\
0 & 1 & 0 & 0 & 0 & 1 \\
1 & 0 & 1 & 0 & 0 & 0 \\
1 & 0 & 0 & 1 & 0 & 0 
\end{array}\right)
\end{equation}

The total coupling matrix for the two maps is therefore:

\begin{equation}
J=\frac 16 \left( 
\begin{array}{c c c c c c}
0 & 1 & 0 & 0 & 1 & 2 \\
1 & 0 & 2 & 1 & 0 & 0 \\
0 & 2 & 0 & 1 & 1 & 0 \\
0 & 1 & 1 & 0 & 1 & 1 \\
1 & 0 & 1 & 1 & 0 & 1 \\
2 & 0 & 0 & 1 & 1 & 0 
\end{array}\right)
\end{equation}

In addition to pyramidal cells, the network contains long-range, inhibitory interneurons whose activity is modeled by a global inhibition on place cells. We assume that the main effect of inhibition is to fix the total neural activity. We introduce the parameter $f$ to denote the fraction of active cells:
\begin{equation}\label{activity}
\sum _{i=1}^N\sigma _i= f\, N \ . 
\end{equation}

Once the coupling matrix $J_{ij}$ (\ref{rule2}) and the constraint over the global activity (\ref{activity}) are defined the probability of a neural activity configuration $\boldsymbol \sigma =(\sigma_1,\sigma_2,\ldots ,\sigma_N)$ is assumed to be
\begin{equation}\label{dist}
P_J(\boldsymbol\sigma ) = \frac 1{Z_J(T)} \; \exp\big(-E_J[\boldsymbol\sigma]/T)\ ,
\end{equation}
where the 'energy' of the configuration reads
\begin{equation}\label{addmodel}
E_J[\boldsymbol\sigma] = -\sum _{i<j} J_{ij} \,\sigma_i\, \sigma_j \ ,
\end{equation}
and the partition function is defined through
\begin{equation}
Z_J (T)= \sum _{\boldsymbol\sigma \ \hbox{\rm with constraint (\ref{activity})}} \exp \big( - E_J[\boldsymbol\sigma] /T) \ .
\end{equation} 
Parameter $T$, which plays the role of temperature in statistical mechanics, fixes the amount of noise in the model. Large values of $T$ corresponds to essentially flat distributions over the neural configuration space. Low $T$ concentrate the probability distribution $P_J$ around the configurations with lowest energies $E_J$.\\
In all numerical computations hereafter we will take the parameters values $w=0.05$ and $f=0.1$, except in Section \ref{parameters} where these values will be discussed.

\subsection{Case of a single environment}

The present model is an extension of the Hopfield model to the case of space-dependent interactions \cite{Hopfield82}. Despite this additional complexity in the model it remains exactly solvable in the infinite $N$ limit due to the extensivity of the number of neighbors of each spin \cite{lebowitz66}.

We start by considering the case of a single environment, for which the coupling matrix is given by (\ref{rule1}). To lighten notations we consider the $D=1$ case; equations for the $D=2$ case are found in Appendix \ref{app2D1env}. In the large $N$ limit, a continuous approach can be introduced by defining the locally coarse-grained activity
\begin{equation}\label{density}
\rho(x) \equiv \lim _{\epsilon \to 0}\lim _{N \to \infty} \; \frac 1{\epsilon N} \sum_{(x-\frac \epsilon 2)N\le i < (x-\frac \epsilon 2)N} \langle \sigma _i \rangle_J \ ,
\end{equation}
where $\langle .\rangle_J$ denotes the average over distribution $P_J$ (\ref{dist}). Due to the presence of periodic boundary conditions we choose $x\in [-\frac 12;\frac 12]$. The density of activity $\rho(x)$ is found upon minimization of the free energy functional
\begin{eqnarray}
&&{\cal F}\big( \{\rho (x)\}\big)=  -\frac 12 \int \mathrm{d}x\, \mathrm{d}y \; \rho(x) J_w(x-y) \rho(y)  \label{functional}\\
&+& T\int  \mathrm{d}x \bigg[ \rho(x) \log \rho(x) + (1-\rho(x)) \log(1-\rho(x)) \bigg]\ , 
\nonumber
\end{eqnarray}  
where $J_w(u)=1$ if $|u|<\frac w2$, and 0 otherwise. The minimum is taken over the activity densities fulfilling
\begin{equation}\label{constraint1}
\int \mathrm{d}x \; \rho(x)=f \ .
\end{equation}
All integrals run over the $[-\frac 12;\frac 12]$ interval.

The minimization equation for $\rho(x)$ can be written as
\begin{eqnarray}\label{extr}
\rho(x) &=& \frac {1}{1+e^{-\mu(x)/T}}\ , \\ 
\mu(x) &=& \int \mathrm{d}y\, J_w(x-y)\rho(y) +\lambda \ , \label{extr2}
\end{eqnarray}
where $\mu(x)$ plays the role of a chemical potential, and the constant $\lambda$ is chosen to satisfy constraint (\ref{constraint1}). We will discuss the different solutions of these equations in the following sections. Note that the free energy per site,
\begin{equation}
F(T) = \lim _{N\to\infty} -\frac TN \log Z_J(T) \ ,
\end{equation}
is simply given by the value of the free-energy functional ${\cal F}$ in its minimum $\rho(x)$, solution of (\ref{extr},\ref{extr2}).

\subsection{Relationship with rate models}
\label{linkwithrate}

Neurons are often described by their firing rate, {\em i.e.} the short-term average of the number of spikes they emit. A straightforward relationship can be drawn with binary models \cite{Ginzburg94}. The current incoming onto neuron $i$ evolves according to
\begin{equation}
\tau \frac{\mathrm{d}I_i}{\mathrm{d}t}=-I_i+\sum_j J_{ij}\; g(I_j) \ .
\end{equation}
Here, $g(x)$ is the characteristic function expressing the firing rate of the neuron as a function of the current. It is a sigmoidal function, running between 0 and 1 (saturation of the postsynaptic neuron at high currents), and $J_{ij}$ includes both the positive coupling $J^0$ (\ref{rule1}) between neighboring cells, and a constant, global inhibition contribution $J^I$, whose value is chosen to enforce condition (\ref{activity}). The dynamical equation admits a stationary state, implicitly defined through
\begin{equation}\label{johnstat}
I_i = \sum_j J_{ij}\; g(I_j) \ .
\end{equation}
Identifying 
\begin{equation}
I_i \to \mu_i \ , \quad g(I_i) \to \rho_i \ ,
\end{equation}
and choosing
\begin{equation}
g(I)= \frac 1{1+\exp(-I/T)} \ ,
\end{equation}
we observe that equation (\ref{johnstat}) for the stationary currents is identical to equation (\ref{extr}) for the chemical potential in the single-environment case. 
The constant term $\lambda$ in (\ref{extr}) is related to the constant inhibitory contribution to $J$ through $\lambda=J^I\,f$. Parameter $T$ fixes the slope of $g$ at the origin. 

As a consequence, the observables of the Ising model (density of activity, chemical potential) are in one-to-one correspondence with the defining features (firing rates, currents) of the stationary states of the neural dynamics. Note that this correspondence also holds in the case of multiple environments, as we shall see in the next Section.

\section{Statistical mechanics of the multiple environment case}\label{multenvt}

\subsection{Average over random remappings}

In the presence of multiple environments the partition function $Z_J$ becomes a stochastic variable, which depends on the $L$ remappings, or, equivalently, on the $L$ random permutations $\pi^\ell$, with $\ell=1 \ldots L$. We assume that, in the large $N$ limit, the free energy of the system is self-averaging, {\em i.e.} concentrated around the average. To compute the average free energy we need to average the logarithm of $Z_J(T)$ over the random permutations. To do so we use the replica method: we first compute the $n^{th}$ moment of $Z_J(T)$, and then send $n\to 0$. The neural configuration is now a set $\vec{\boldsymbol \sigma} = (\boldsymbol\sigma^1, \ldots, \boldsymbol\sigma^n)$ of $n\times N$ spins $\sigma_i^a$, where $i=1...N$ is the spin index and $a=1\ldots n$ is the replica index. The $n^{th}$ moment of the partition function reads

\begin{eqnarray}
\overline{Z_J(T)^n} &=& \sum _{\vec{\boldsymbol\sigma}} \overline{\exp\left[ \beta \sum_{a=1}^n \sum_{i<j} \left( J^0_{ij}+ \sum_{\ell =1}^L J_{ij}^\ell \right) \,\sigma_i^a \sigma_j ^a\right] }\nonumber \\
&=& \sum _{\vec{\boldsymbol\sigma}} \exp\left[\beta  \sum_{a=1}^n \sum_{i<j} J^0_{ij} \,\sigma_i^a \sigma_j ^a\right] \; \Xi \big(\vec{ \boldsymbol\sigma}\big)^L
\ ,
\end{eqnarray}
where $\beta=1/T$ and the overbar denotes the average over the random remappings. The sum over $\vec{\boldsymbol\sigma}$ is restricted to configurations with average activity equal to $f$ (within each replica), and
\begin{equation}
\Xi \big(\vec{\boldsymbol \sigma}\big) = \frac 1{N!} \sum _{\pi^\ell} \exp\left[ \beta  \sum_{i<j} J^0_{ij} \sum_{a=1}^n \sigma_{\pi^\ell(i)}^a \sigma_{\pi^\ell(j)} ^a\right] \ .
\end{equation}
The calculation of the average over the random permutation $\pi^\ell$ is not immediate, but can be done exactly in the large $N$ limit. Details are reported in Appendix \ref{appaverage}. The result is 
\begin{eqnarray}\label{expxi}
\log \Xi \big(\vec{\boldsymbol \sigma}\big)  &=& -\frac\beta 2nf(1-f) +N \frac \beta 2n w f^2 \\
&-& \sum_{\lambda \ne 0} \text{Trace} \;\log\left[\text{\bf Id}_n - \beta \lambda \big({\bf q}-f^2\, {\bf 1}_n\big)\right]  \nonumber
\ ,
\end{eqnarray}
where {\bf Id}$_n$ denotes the $n$--dimensional identity matrix, ${\bf q}$ is the overlap matrix with entries 
\begin{equation}\label{defqab}
q^{ab} \equiv \frac 1N \sum_{j} \sigma_j^{a} \sigma_j^{b} \ ,
\end{equation}
and {\bf 1}$_n$ is the $n\times n$ matrix whose all entries are equal to one. The sum in (\ref{expxi}) runs over all the non-zero eigenvalues of the matrix $J^0$. Explicit expressions for those eigenvalues will be given in the next Section for the $D=1$ case, while the two-dimensional case is treated in Appendix \ref{app2D}.

A key feature of expression (\ref{expxi}) is that $\Xi$ depends on the spin configuration $\vec{\boldsymbol\sigma}$ through the overlaps $q^{ab}$ only. Those overlaps thus play the role of order parameters for the activity in the environment $\ell \ge 1$, as does $\rho(x)$ for the environment 0. Calculation of the $n^{th}$ moment of the partition function therefore amounts to estimating the entropy of neural acitvity configuration $\vec{\boldsymbol\sigma}$ at fixed $\{q^{ab},\rho(x)\}$, which can be done exactly in the $N\to\infty$ limit.

\subsection{Replica-symmetric theory}

To perform the $n\to 0$ limit we make use of the replica symmetric Ansatz, which assumes that the overlaps $q^{ab}$ take a single value, $q$, for replica indices $a\ne b$. The validity of the Ansatz will be discussed in Section \ref{sectrois}. The Edwards-Anderson order parameter $q$, defined through
\begin{equation}
q \equiv \frac 1N \sum _{i=1}^N \overline{\langle \sigma_i \rangle_J ^2} \ ,
\end{equation}
measures the fluctuations of the local spin magnetizations from site to site. Values for $q$ range from $f^2$ to $f$. We expect $q$ to be equal to $f^2$  when the local activity $\langle\sigma_i\rangle_J$ (averaged over the configurations with distribution $P_J$)  is uniform over space, and to be larger otherwise.

As in the single environment case we define the order parameter $\rho(x)$ as the density of activity around point $x$ in space, see (\ref{density}),
\begin{equation}\label{density_ave}
\rho(x) \equiv\lim _{\epsilon \to 0}\lim _{N \to \infty} \; \frac 1{\epsilon N} \sum_{(x-\frac \epsilon 2)N\le i < (x-\frac \epsilon 2)N}\overline{ \langle \sigma _i \rangle_J }\ ,
\end{equation}
The difference is that, in the multiple environment case, the density $\rho(x)$ appearing in the replica theory is averaged over the environments. Local fluctuations of the density from environment to environment can be calculated \cite{Monasson13}, but will not be considered here; only global fluctuations, averaged over space, are considered through the order parameter $q$.

As in the single environment case  a chemical potential $\mu(x)$, conjugated to $\rho (x)$,  is introduced. In addition, a new order parameter, $r$, is necessary to describe the force conjugated to $q$, and controlling the fluctuations of the spin magnetizations. All order parameters are determined through the optimization of the free-energy functional ${\cal F}(q,r,\{\rho(x)\},\{\mu(x)\})$, see Appendix \ref{appreplica}, whose expression for the $D=1$ case is given by
\begin{eqnarray}\label{calf}
{\cal F} &=& \frac {\alpha \beta}2 r(f-q) -\frac\alpha\beta\psi(q,\beta)+ \int \mathrm{d}x\, \mu(x)\,\rho(x) \nonumber \\
 &-&\frac 12 \int \mathrm{d}x \int \mathrm{d}y \,\rho(x)\,J_w(x-y)\,\rho(y) \\
&-&\frac 1\beta\int \mathrm{d}x \int Dz \log \bigg( 1 + e^{\beta
z\sqrt{\alpha r} + \beta\mu(x)}\bigg)\nonumber \ ,
\end{eqnarray}
where $Dz=\exp(-z^2/2)/\sqrt{2\pi}$ is the Gaussian measure, and
\begin{eqnarray}
\psi(q,\beta)& \equiv&\sum\limits_{k\geq 1}\left[\frac{\beta(q-f^2)\sin(k\pi w)}{k\pi- \beta(f-q)\sin(k\pi w)}\right. \nonumber \\
&-&\left.\log\left(1-\frac{\beta(f-q)\sin(k\pi w)}{k\pi}\right)\right]\ .
\end{eqnarray}
Parameter $\alpha\equiv L/N$, hereafter called load, denotes the ratio of the numbers of environments and of cells. Again, optimization is done over densities $\rho(x)$ fulfilling constraint (\ref{constraint1}). 

Extremization of the free energy functional leads to the saddle-point equations 
\begin{eqnarray}\label{eq:saddlepoint1}
r&=&2(q-f^2)\sum\limits_{k\geq 1}\left[\frac{k \pi}{\sin(k\pi w)}-\beta(f-q)\right]^{-2}\nonumber \ ,\\
q&=&\int\mathrm{d}x\int\mathrm{D}z\; \big[1+e^{-\beta z\sqrt{\alpha r}-\beta\mu(x)}\big]^{-2}\ ,\nonumber\\
\rho (x)&=&\int\mathrm{D}z\; \big[1+e^{-\beta z\sqrt{\alpha r}-\beta\mu(x)}\big]^{-1}\ ,\\
\mu(x)&=&\int\mathrm{d}y\, J_w(x-y)\, \rho(y)+\lambda\ ,\nonumber
\end{eqnarray}
where $\lambda$ is determined to enforce constraint (\ref{constraint1}).

The expression of ${\cal F}$ and of the saddle-point equations for the $D=2$ case can be found in Appendix \ref{app2D}.

\section{The phases and their stability}\label{sectrois}

In both $D=1$ and 2 dimensions three qualitatively different solutions are found for the extremization equations of ${\cal F}$, corresponding to three distinct phases of activity: a paramagnetic phase in which the activity is uniform over space, a 'clump'-like phase in which the activity is localized in one of the stored spatial maps, and a glassy phase where the activity is neither uniform nor coherent with any map. We now discuss the domains of existence and stability of each phase. We are chiefly interested in the clump phase domain, which corresponds to the experimentally observed regime where memorized maps can be retrieved. As usual all expressions given below correspond to the $D=1$ case, while the case $D=2$ is treated in Appendix \ref{app2D}; all numerical results will be given taking $f=0.1$, $w=0.05$.

\subsection{High noise: Paramagnetic phase}

\begin{figure}
\begin{center}
\includegraphics[width=8.5cm]{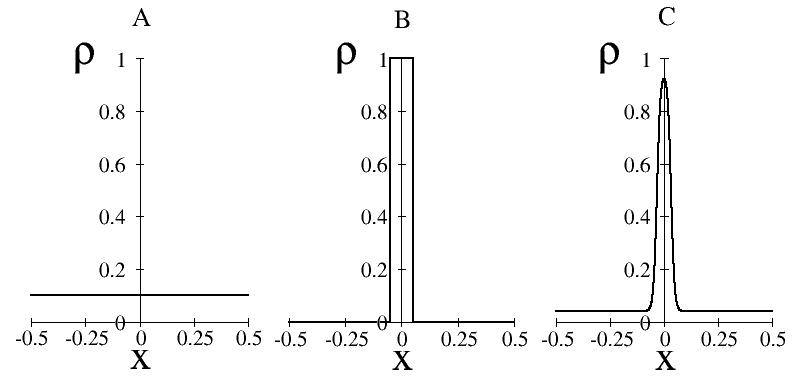}
\caption{Average activity $\rho(x)$ in dimension $D=1$ in the paramagnetic phase ({\bf A}) and in the clump phase ({\bf B}: temperature $T=0$, {\bf C}: temperature $T=0.0073$) for $\alpha=0$, computed with $M=2000$ bins of discretization. }
\label{activity_clump}
\end{center}
\end{figure}

At high temperature we expect the activity to be dominated by the noise in the neural dynamics, and to show no spatial localization. The corresponding order parameters are:
\begin{equation}
\rho (x) =f \ , \quad q=f^2\quad \text{(paramagnetic phase - PM)} \ .\nonumber
\end{equation}
The activity profile is shown in Fig.~\ref{activity_clump}A. The paramagnetic phase (PM) exists for all values of the control parameters, with corresponding potentials:
\begin{equation}
\mu(x)+\lambda =T\log\left(\frac{f}{1-f}\right) \ , \quad r =0\quad \text{(PM)} \ .\nonumber
\end{equation}
We now discuss its stability.

\subsubsection{Case of a single environment ($\alpha = 0$)}

In the single environment case the stability of the paramagnetic solution is determined by computing the Hessian of the free-energy  functional ${\cal F}$ (\ref{functional}). We find that
\begin{equation}
\frac{\delta^2{\cal F}}{\delta \rho(x)\delta \rho(y)} = \frac T{f(1-f)}\delta(x-y)- J_w(x-y) \label{functional2} \ .
\end{equation}  
The solution is stable as long as the Hessian is definite positive. 

In the one-dimensional case the most unstable mode corresponds to a spin wave $\delta \rho(x) \propto \sin(2\pi\,k\, x)$, with wave number $k=1$; note that the $k=0$ mode is forbidden according to  condition (\ref{constraint1}). The unstability develops under the spinodal temperature
\begin{equation}\label{deftpm}
T_{PM} = f(1-f)\;\frac{\sin \pi w}\pi\approx0.0045 \ .
\end{equation}
Note that $T_{PM}$ and, more generally, all thermodynamic quantities are invariant under the changes $f\to 1-f$ or/and $w\to 1-w$, which simply amount to reverse ${\sigma_i\to 1-\sigma_i}$, {\em i.e.} to change active spins into holes and vice versa.

\subsubsection{Case of multiple environments ($\alpha >0$) }

\begin{figure}
\begin{center}
\includegraphics[width=\linewidth]{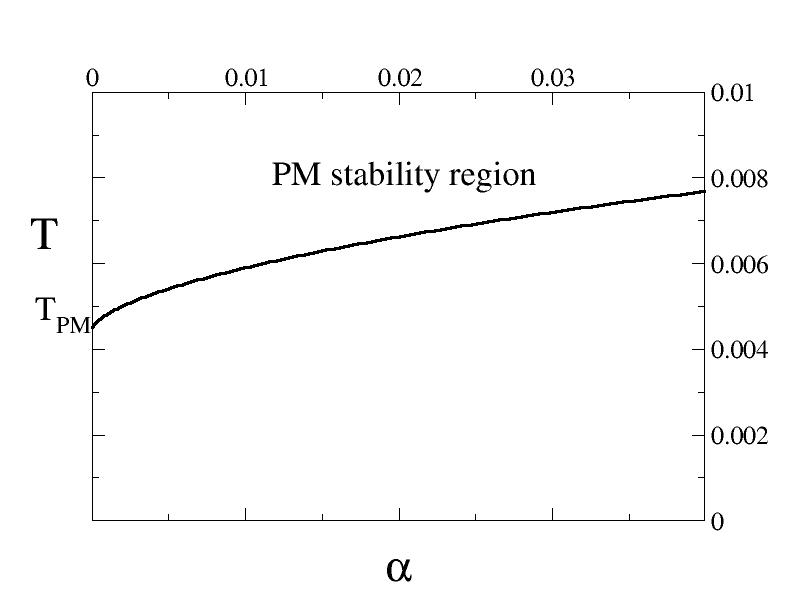}
\caption{Paramagnetic phase stability region in the $(T,\alpha)$ plane, defined by $T>T_{PM}(\alpha)$.}
\label{fig:stabpara}
\end{center}
\end{figure}

The study of the stability of the PM phase in the multiple environments case is reported in Appendix \ref{appstabpara}. As in the single environment case the PM solution is unstable at all temperatures $T<T_{PM}$ against perturbation of the activity of the type $\delta \rho(x) \propto \sin(2\pi\,k\, x)$. In addition coupled fluctuations of $\lambda,r,q$ may lead to instabilities if $T$ is smaller than ${T}_{PM}(\alpha)$, implicitly defined through
\begin{equation}
\sum\limits_{k\geq1}\left[\frac{{T}_{PM}(\alpha)\,k \pi}{f(1-f)\sin(k\pi w)}-1\right]^{-2}=\frac{1}{2\alpha} \ .
\end{equation}
 The instabilities correspond to the transition to the glassy phase, see Section \ref{secglass}.  
Note that $T_{PM}$ defined in (\ref{deftpm}) corresponds with $T_{PM}(\alpha=0)$. As a conclusion, in the $(\alpha,T)$ plane, the PM phase is stable in the region $T>T_{PM}(\alpha)$. This region is sketched in Fig.~\ref{fig:stabpara}. 

\subsection{Moderate noise and load: The clump phase}

In experiments place cells exhibit patterns of localized activity where neurons with neighboring place fields are active together. Our modelling reproduces such localized--in--space activity patterns (called 'bumps' or 'clumps' of activity) at sufficiently low ($\alpha,\ T$). The corresponding phase, hereafter referred to as 'clump phase' (CL), is characterized by the order parameters:
\begin{equation}
\rho(x) \ne f \ , \quad q > f^2 \quad \text{(clump phase - CL)}\ . \nonumber
\end{equation}
Correspondingly, the chemical potential $\mu(x)$ will vary over space, and the conjugated force $r$ is strictly positive.

\subsubsection{Case of a single environment ($\alpha=0$)}
\label{clump_1env}
When the temperature $T$ is sent to $0^+$, assuming that $f>w$ we find a solution to (\ref{extr}) that is localized in space:
\begin{equation}\label{mu1d}
\mu(x) = \left\{ \begin{array} {c c c} 
w & \hbox{\rm if} & |x| < \frac 12 (f-w) \\
\frac 12 (f+w) - |x| & \hbox{\rm if} & \frac 12 (f-w) \le |x| < \frac 12 (f+w) \\
0 & \hbox{\rm if} & |x| \ge \frac 12 (f+w)
\end{array} \right. \ , 
\end{equation}
and
\begin{equation}
\rho(x) \to \left\{ \begin{array} {c c c} 
1 & \hbox{\rm if} & |x|\le f/2 \\
0 & \hbox{\rm if} & |x| >f/2
\end{array} \right. \ .
\end{equation}
Any translation $x\to x+x_0$, with $x_0 \in ]0;1[$, defines another ground state with the same energy. The activity profile is shown in Fig.~\ref{activity_clump}B.

At small but finite temperature we have solved equations (\ref{extr}) numerically by discretizing  space with a large number $M$ of bins of width $1/M$, such that $Mw$ and $Mf$ are both much larger than unity.  

\begin{figure}
\includegraphics[width=\linewidth]{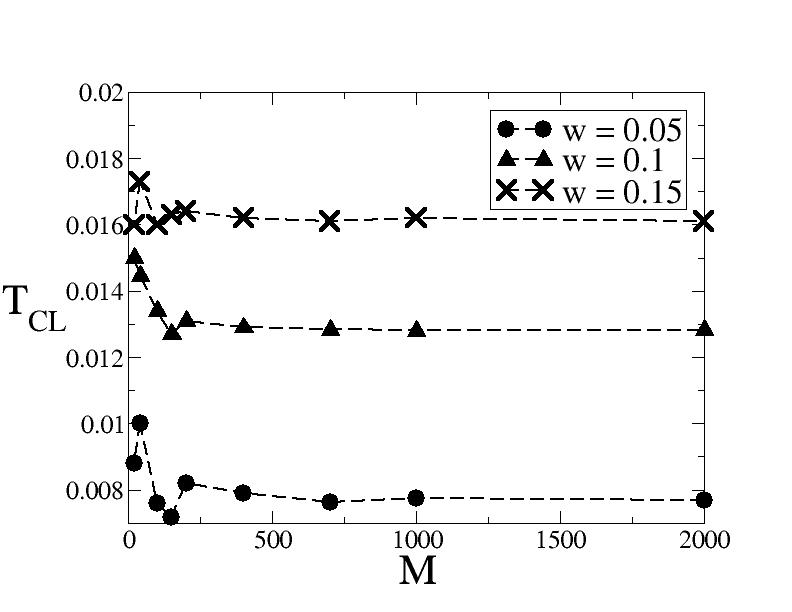}
\caption{Numerical effect: $T_{CL}$ as a function of the number $M$ of discretization bins for $w=0.05$ (circles), $w=0.1$ (triangles) and $w=0.15$. $T_{CL}$ reaches its asymptotic value at a saturation $M$ that is a decreasing function of $w$.}
\label{fig:bins}
\end{figure}

The shape of the clump of activity is now rounded off by the thermal noise; in addition, far away cells are active with some positive probability $<f$. This clump is reminiscent of a liquid phase, surrounded by its vapor. The clump persists up to some critical temperature $T_{CL}$, at which it disappears. The value of $T_{CL}$ depends on $f$ and $w$, {\em e.g.} $T_{CL}\simeq 0.008$ for $f=0.1,w=0.05$. A representative shape of the activity clump at finite temperature is shown in Fig.~\ref{activity_clump}C. The dependency on $f$ and $w$ will be studied in section \ref{discussion}. Notice that $T_{CL}$ also slightly depends on the number of bins of discretization $M$ as shown in Fig.~\ref{fig:bins}. 

\begin{figure}
\includegraphics[width=\linewidth]{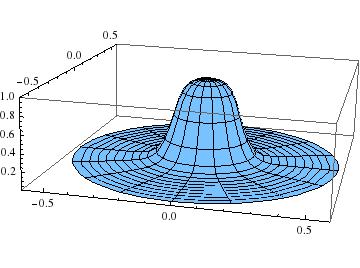}
\includegraphics[width=\linewidth]{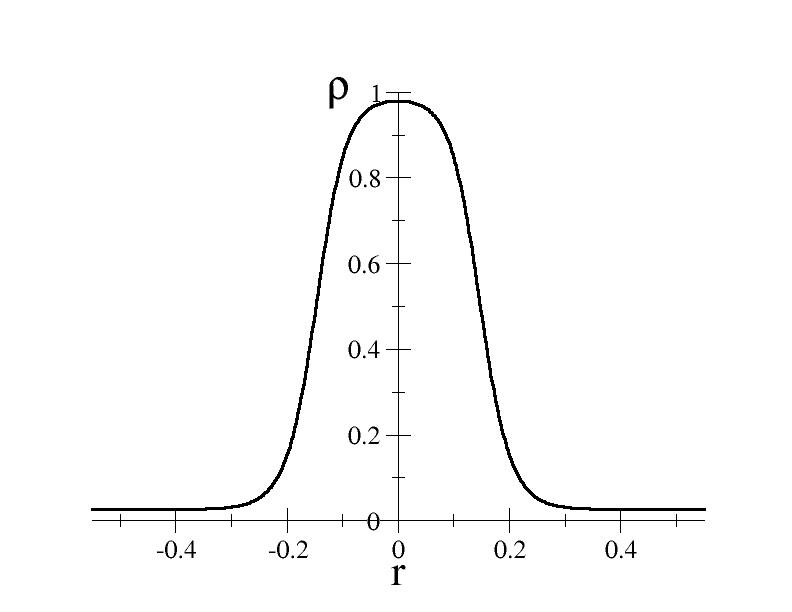}
\caption{Two-dimensional clumps of activity for a single environment ($\alpha=0$) at temperature $T=0.0055$ computed with $M=400$. The whole clump $\rho(x,y)$ is shown in the top panel, while the bottom panel shows the radial cut of the profile.}
\label{fig:clump2D}
\end{figure}

The clump phase is also found to be present in dimension $D=2$. At finite temperature we solve equations (\ref{eq:saddlepoint2d}) numerically as in the one-dimensional case. An example of two-dimensional clump is shown in Fig.~\ref{fig:clump2D}. 

\subsubsection{Case of multiple environments ($\alpha >0$)}

\begin{figure}
\begin{center}
  \includegraphics[width=\linewidth]{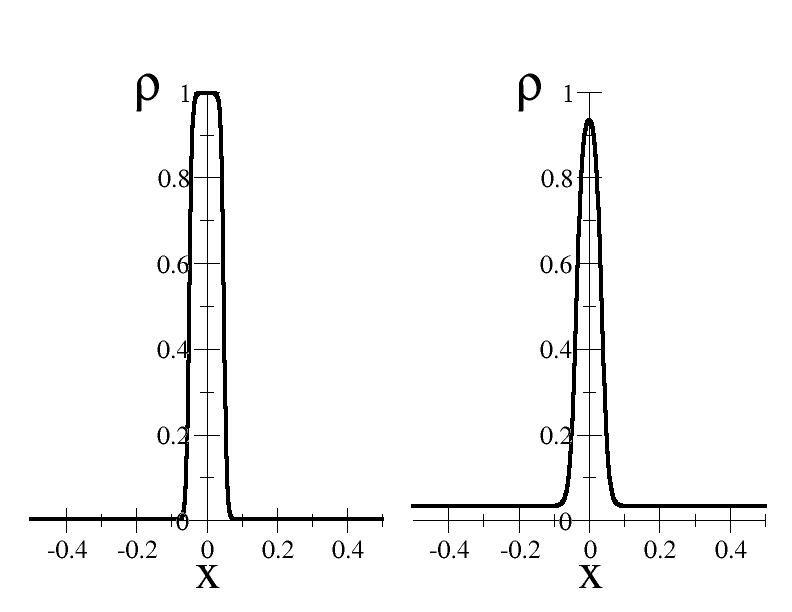}
\caption{Effect of the load $\alpha$ on the clump: average activity $\rho(x)$ in dimension $D=1$ in the clump phase at temperature $T=0.004$ for $\alpha=0$ (left) and $\alpha=0.02$ (right).}
\label{fig:clumpdea}
\end{center}
\end{figure}

We look for a solution with localized activity in the first environment, and non-localized activity in the other environments. We have solved the coupled extremization (\ref{eq:saddlepoint1}) in one dimension using the numerical procedure described above. We observe that, at a given temperature, increasing $\alpha$ has the effect of squeezing and lowering the clump (Fig.~\ref{fig:clumpdea}). Note that, because the disorder is averaged, the clump solution is invariant by translation in space as in the single environment case (bumps of activity centered on all positions have the same free-energy). Nevertheless, for a given realization of disorder this invariance by translation does not exist. In any case, when the rodent wanders in a familiar environment, some input containing information about its position (sensory cues and/or path integration) is supposed to act as a strong local field that selects the clump centered on the right position. Here we consider the states of the network in the absence of external input.

\begin{figure}
\begin{center}
\includegraphics[width=\linewidth]{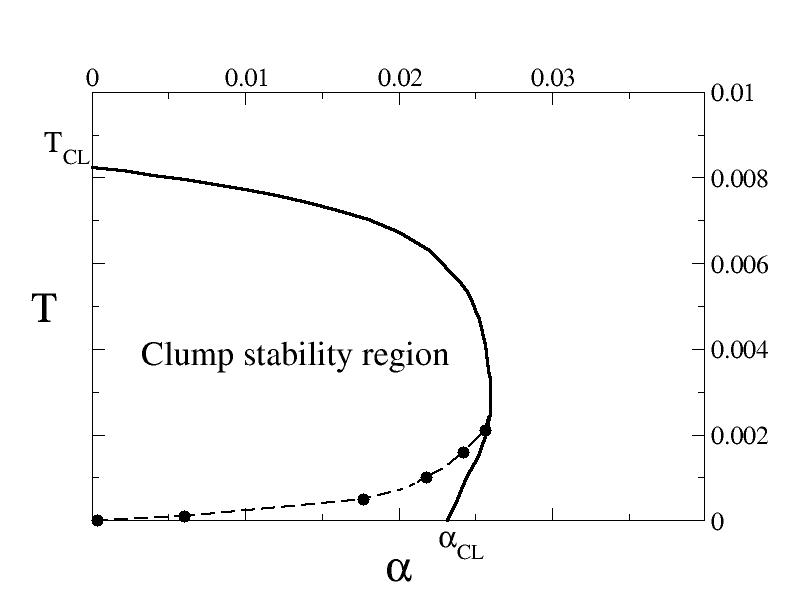}
\caption{Domain of stability the clump phase, computed with $M=200$. The longitudinal and replicon instability lines correspond to, respectively, the full and dashed lines. Because of numerical calculation times, only a few points of the replicon line could be computed; they are represented by circles.}
\label{fig:stabclump}
\end{center}
\end{figure}

We have studied the stability of the clump solution against longitudinal and replicon modes. The longitudinal stability domain is found by  determining the boundary in the $(\alpha,T)$ along which the clump abruptly collapses. This boundary, shown in Fig.~\ref{fig:stabclump}, can be described by two sections of curves: 
\begin{itemize}
\item at small $\alpha$ the clump phase is longitudinally stable for $T< T_{CL}(\alpha)$, a slowly decreasing function of $\alpha$, which coincides with the temperature $T_{CL}$ found for a single environment when $\alpha \to 0$. 
\item at small temperature, the clump phase is longitudinally stable if $\alpha<\alpha_{CL}(T)$, an increasing function of $T$. We denote $\alpha_{CL}$ its value when $T\to 0$.
\item At intermediate temperatures a weak reentrance is present. The curves $T_{CL}(\alpha)$ and $\alpha_{CL}(T)$ merge at a point where the tangent is vertical and the reentrance begins.
\end{itemize}
Numerically, a slight dependency on $M$ is observed.

Along the boundary of the clump phase the value of the Edwards-Anderson parameter increases from $q=f^2$ in $(\alpha=0,T=T_{CL})$ to $q=f$ in $(\alpha=\alpha_{CL},T=0)$.

Calculation of the stability against replicon modes is detailed in Appendix \ref{appstabclump}. We find that the replica-symmetric solution is stable, except in a small region confined to small $T$ and $\alpha$ close to $\alpha_{CL}$. This result is shown by the dashed line in Fig.~\ref{fig:stabclump}. It is reminiscent of the results for the 'retrieval phase' in the Hopfield model \cite{Amit85}.

\subsection{High load: The glassy phase}
\label{secglass}

At large $\alpha$ the disorder in the interactions is strong enough to magnetize the spins locally, without any coherence with any spatial map. Again, the average of the activity $\langle \sigma_i\rangle_J$ will depend on the realization of the environments, while the average over the environment,
$\overline{\langle \sigma_i\rangle_J}$ will be uniform in space and equal to $f$. In this glassy (SG) phase the order parameters will take values
\begin{equation}
\rho(x)=f \ , \quad q > f^2 \quad \text{(glassy phase - SG)}\ .\nonumber
\end{equation}
Correspondingly the chemical potential $\mu(x)$ does not depend on $x$, and $r>0$.

\begin{figure}
\begin{center}
\includegraphics[width=\linewidth]{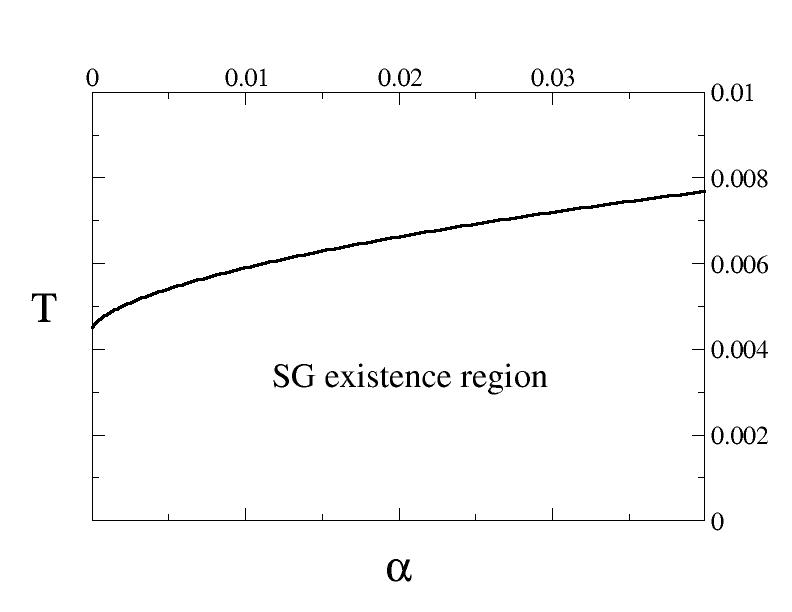}
\caption{Domain of existence of the glassy phase in the $(T,\alpha)$ plane. The phase is always stable against longitudinal stability. Replica symmetry is always broken.}
\label{fig:stabglass2}
\end{center}
\end{figure}

As reported in Appendix \ref{appstabglass} the glass phase is found when $T<T_{SG}(\alpha)$, where $T_{SG}(\alpha)$ is the same line $T_{PM}(\alpha)$ found above. Within this region, the SG phase is always stable against clumpiness (localization of the activity). The spin glass phase is always unstable against replicon mode, indicating that replica symmetry is always broken, similarly to the spin glass phase in the Hopfield model \cite{Amit85}.  Results are summarized in Fig.~\ref{fig:stabglass2}.

\section{Phase diagram}
\label{phase_diagram}

\subsection{Transitions between phases}

\begin{figure}
\begin{center}
\includegraphics[width=\linewidth]{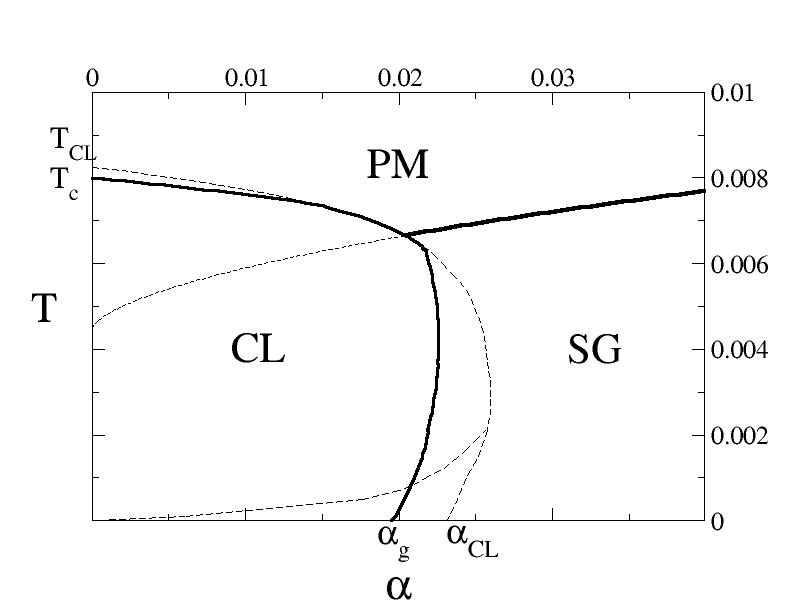}
\caption{Phase diagram in the $(T,\alpha)$ plane in $D=1$. Thick lines: transition between phases. Thin dashed lines reproduce stability regions described above. Critical lines are computed with $M=200$.}
\label{fig:phasediag}
\end{center}
\end{figure}

The first-order phase transitions occur when two phases have the same free energy. The critical lines are found numerically. In dimension 1,
\begin{itemize}
 \item The clump-paramagnetic transition at high temperature occurs slightly before the clump unstability line. We denote $T_c(\alpha)$ the corresponding temperature for a given $\alpha$.
\item The clump-glass transition occurs at a load denoted $\alpha_g(T)$ for a given temperature $T$. Here again, we find a slight reentrance at moderate temperature: $\alpha_g(T)$ is maximal for $T\approx0.004$. Since the glassy phase has been shown to be replica-symmetry broken, its free energy is expected to be higher than in the RS case; therefore the 'real' transition is expected to be slightly shifted to higher values of $\alpha$. 
\item At high $\alpha$, $T$ there is a second-order transition between the PM and the SG phases.
\end{itemize}
The phase diagram in dimension 1 is summarized in Fig.~\ref{fig:phasediag}.

It is interesting to emphasize the differences between this phase diagram and the one of the Hopfield model computed in \cite{Amit85}. In the Hopfield model, the 'retrieval' or 'ferromagnetic' (FM) phase (which corresponds to our clump phase) has a triangular shape in the $(\alpha,T)$ plane. The temperature at which the FM phase becomes unstable at a given $\alpha$ is smaller than $T_{PM}(\alpha)$. There is no coexistence between the PM and FM phases, and both are separated by the glassy phase. Moreover, for the Hopfield model, $T_{FM}(\alpha)$ is monotonously decreasing so the capacity is maximal at zero temperature \footnote{A slight reentrance was found in the RS solution in later works \cite{Canning92, Steffan94}, but it is very weak.}. Consequently, it seems that our model of attractor neural network is much more robust to noise than the standard Hopfield model. This can be understood considering the structure of the coupling matrix. In the Hopfield model one patterns defines a single direction in the configuration space; interference with other patterns and dynamical noise may push the activity configuration in the high-dimensional orthogonal subspace, and the memory of the pattern is easily lost.  In the present case, on the contrary, one map defines a whole collection of configurations (bumps) centered on different locations, thus the synaptic matrix will make the network converge to one of the attractors, even in the presence of a high level of noise. This robustness to noise will also be an interesting feature in the study of the dynamics of the model.

\begin{figure}
\centering
  \includegraphics[width=\linewidth]{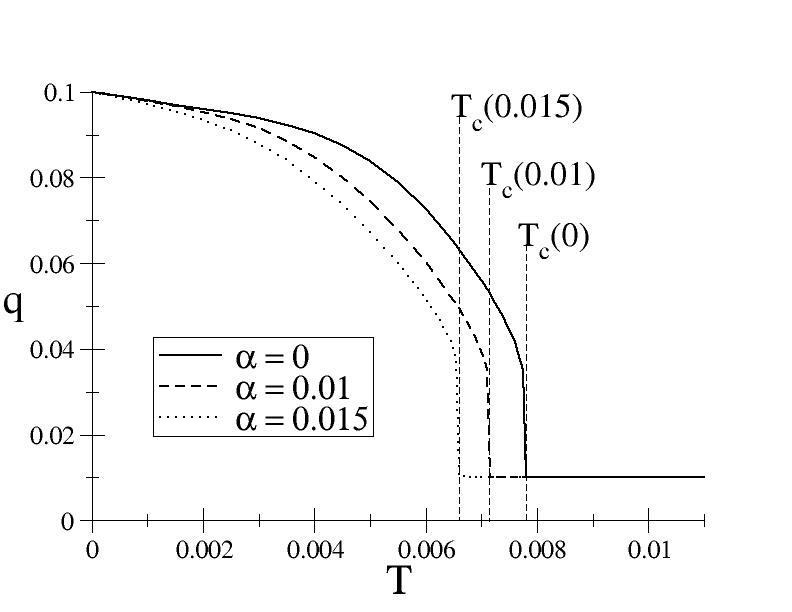}
  \caption{$q$ as a function of $T$ for fixed $\alpha$: {$\alpha=0$} (solid line), {$\alpha=0.01$} (dashed line) and {$\alpha=0.015$} (dots), computed with $M=1000$. A discontinuity is observed at the clump-paramagnetic transition.}
  \label{fig:qjumpa}
  \includegraphics[width=\linewidth]{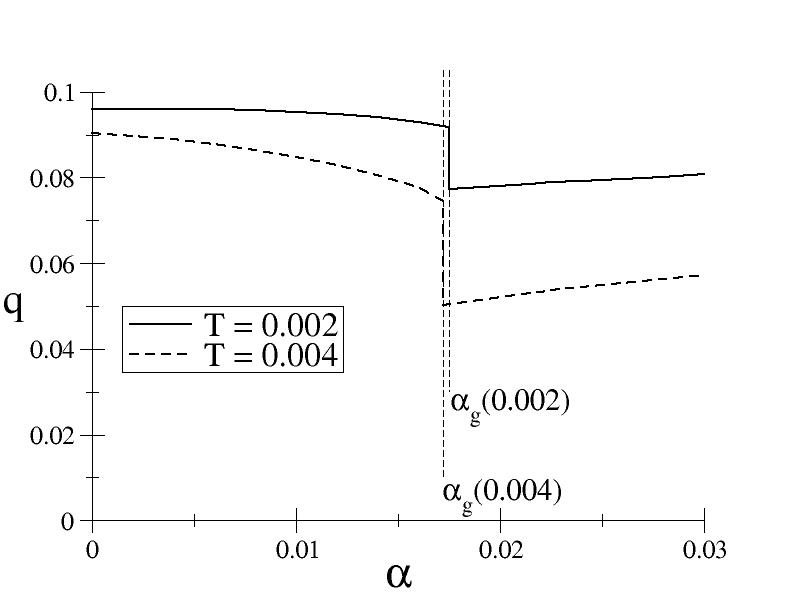}
  \caption{$q$ as a function of $\alpha$ for fixed temperature: ${T=0.002}$ (solid line) and ${T=0.004}$ (dashed line), computed with ${M=1000}$. A discontinuity is observed at the clump-glass transition.}
\label{fig:qjumpb}

\end{figure}

When the transition line is crossed there is a discontinuity in the order parameter $q$. We have computed numerically the value of the Edwards-Anderson parameter at different points and plotted its evolution at the clump-paramagnetic transition at fixed $\alpha$ (Fig.~\ref{fig:qjumpa}) and at the clump-glass transition at fixed $T$ (Fig.~\ref{fig:qjumpb}).

\subsection{Numerical simulations}
\label{descriptionMC}
We have performed Monte Carlo simulations to thermalize the Ising model. The system is initialized with two types of conditions (respectively, uniform and clump configurations). At each time step, two neuron indices $i,j$ are chosen such that $\sigma_i=1-\sigma_j$. We then calculate the change in the energy when the two spins are flipped, and perform the flip or not according to Metropolis' rule. As a consequence the activity is kept constant (and equal to $fN$ over the neural population), and the system is guaranteed to reach equilibrium for sufficiently long simulation times.

\subsubsection{Single environment case}
\begin{figure}
\begin{center}
\includegraphics[width=\linewidth]{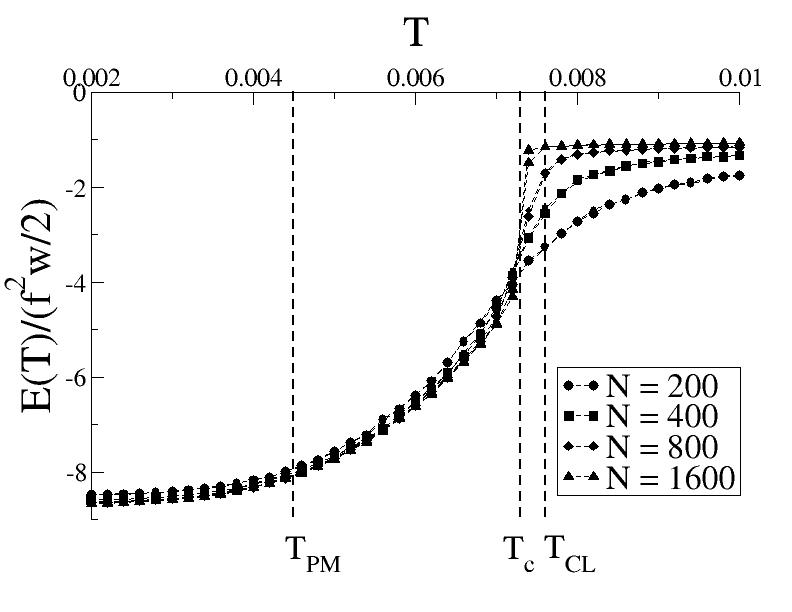}
\caption{Average energy for the unidimensional model with a single environment and for increasing sizes $N$. For each size, we plot the average energy obtained after thermalization for $10\, N$ Monte Carlo steps starting from the uniform and from the clump configurations. Each point is averaged over 1000 simulations.}
\label{energy1env}
\end{center}
\end{figure}

Fig.~\ref{energy1env} shows the average energy $E(T)$ vs. the temperature $T$, for various sizes $N$. At high temperature, $E(T)=-\frac 12f^2 w$ as expected in the paramagnetic phase. At low temperature, the shape of the activity clump varies with $T$, and so does $E(T)$. We find a clear signature of the first order transition as $N$ grows. The critical temperature is in excellent agreement with the analytical value for $T_c$. 

\begin{figure}
\begin{center}
\includegraphics[width=\linewidth]{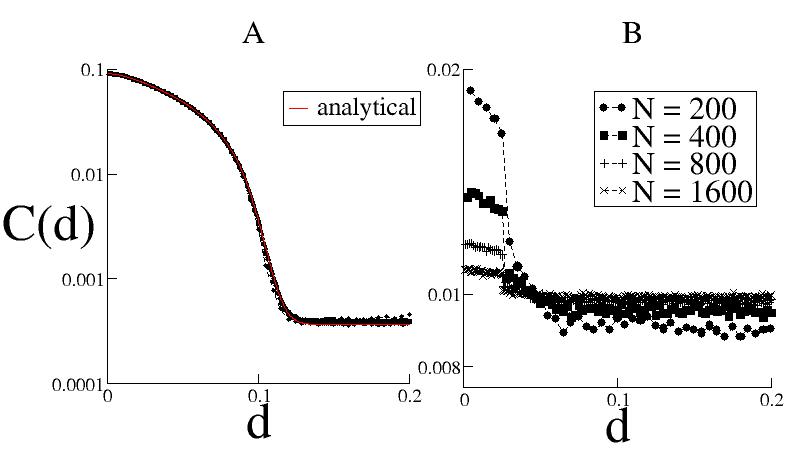}
\caption{Correlation $C(d)$ between spins at distance $d$ (\ref{defcd}) at low (left) and high (right) temperatures, and for various sizes $N$. {\bf A}. $T=.004$, {\bf B}. $T=.01$. Note the difference of logarithmic scale on the y-axis between the two panels. }
\label{correl1env}
\end{center}
\end{figure}

We plot in Fig.~\ref{correl1env} the spin-spin correlation, $\langle \sigma_i \sigma_{j}\rangle$ as a function of the normalized distance, $d=\frac{|i-j|}N$:
\begin{equation}\label{defcd}
C(d)=\langle \sigma _i \sigma_{i+d\,N}\rangle \ .
\end{equation} 
At low temperature, finite size effects are negligible and $C(d)$ is a non trivial decreasing function of $d$ in the large $N$ limit. At small $d$, $C(d)$ is of the order of $f$, and then decreases to a much smaller value over a distance of the order of $f$. As the location of the clump is arbitrary, we expect its center $x_0$ to be uniformly distributed over the $[-\frac 12;\frac 12]$ interval. The correlation is therefore given, in the thermodynamic limit, by
\begin{equation}\label{corrcorr}
C(d) = \int \mathrm{d}x_0 \, \rho(x_0)\, \rho(x_0+d) \ .
\end{equation}
At zero temperature, this formula gives $C(d)=f-d$ for $d<f$, $C(d)=0$ for $d\ge f$. At finite temperature, we compute $\rho$ from the extremization equation (\ref{extr}), and plug the value into the r.h.s. of (\ref{corrcorr}). The agreement with the correlation $C(d)$ obtained from MC simulations is perfect (Fig.~\ref{correl1env}).

At high temperature and for finite $N$, $C(d)$ decreases over a distance $d\simeq \frac w2$ to the paramagnetic value $f^2$. When $N\to\infty$, $C(d)$ is uniformly equal to $f^2$ at all distances $d>0$. As an additional check of the value of $T_c$ we find that the spin-spin correlation decays quickly with increasing $N$ for $T=.0074$, and saturates to a $d$-dependent value larger than $f^2$ for $T=.0072$ (not shown). 

\subsubsection{Multiple environments}

\begin{figure}
\begin{center}
\includegraphics[width=\linewidth]{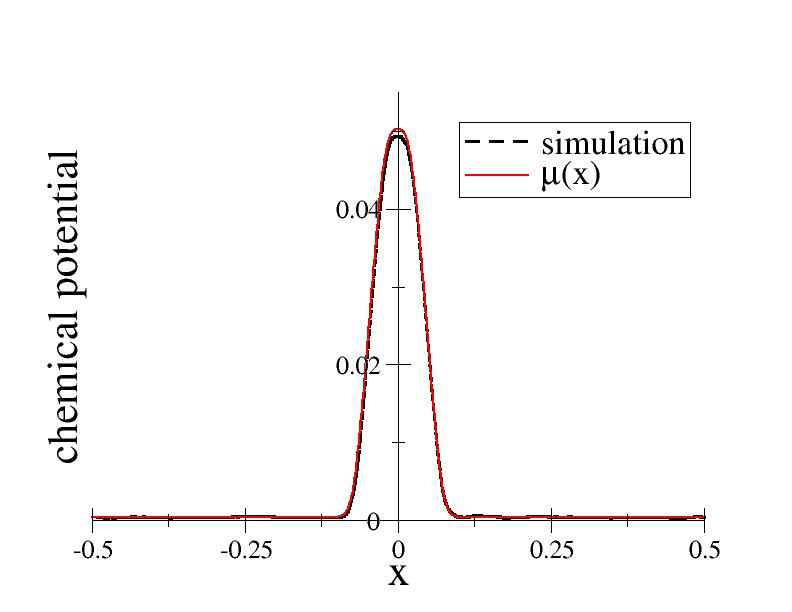}
\caption{Chemical potential $\frac 1N\sum_j J_{ij}^0\sigma_j$ as a function of $x$ for ${\alpha=0.01}$ and $T=0.004$: analytical prediction $\mu(x)$ (red solid line) and result of simulation (black dashed line) with $N=10^4$, averaged on $10^2$ rounds of $10N$ steps.}
\label{fig:mu}
\end{center}
\end{figure}

\begin{figure}
\begin{center}
\includegraphics[width=\linewidth]{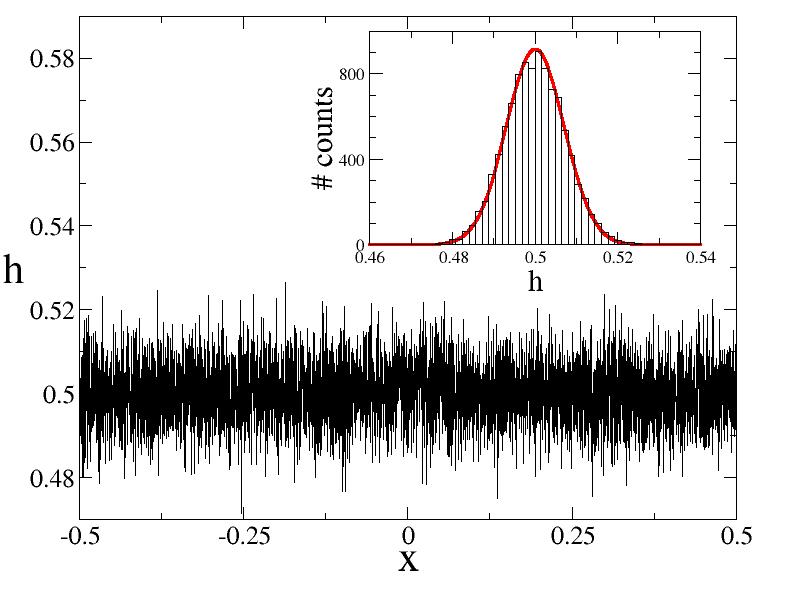}
\caption{Contribution $h_i$ of environments $\ell\geq1$ to the local field as a function of $x$ in simulation with $N=10^4$, averaged on $10^2$ rounds of $10N$ steps. Inset: histogram of $h_i$ (black rectangles) compared to the Gaussian distribution of mean $fLw$ and standard deviation $\sqrt{\alpha r}$ (red line). The value $\sqrt{\alpha r}\simeq6.98\cdot10^{-3}$ is computed from saddle point equations (\ref{eq:saddlepoint1}).}
\label{fig:r}
\end{center}
\end{figure}

The same Monte-Carlo simulations have been performed with several environments, corresponding to random permutations of the sites, additively encoded in the coupling matrix. We have verified numerically the theoretical predictions for $\mu(x)$ (Fig.~\ref{fig:mu}) and $r$ (Fig.~\ref{fig:r}). This latter quantity can be accessed by measuring the local fields at different positions, ${\mu(x)+\lambda+z\sqrt{\alpha r}}$. The quenched noise on the field comes from the contribution of environments $\ell\geq1$: ${ z\sqrt{\alpha r}}$  is a Gaussian random variable of mean 0 and standard deviation $\sqrt{\alpha r}$ independent of $x$. In our simulations we have measured the contribution ${h_i\equiv\frac 1N\sum\limits_{\ell=1}^L\sum\limits_{j}J_{ij}^{\ell}\sigma_j}$   of environments $\ell\geq1$ to the local field at different locations. We have checked that their distribution matched the prediction of a Gaussian of width $\alpha\, r$ with excellent agreement (see inset in Fig.~\ref{fig:r}).


We have also investigated the behavior of the system for varying levels of noise and load, and compared it to the phase diagram found analytically. In simulations we have considered the environment $\ell$ of lowest energy (in which the activity acquires a clump-like shape)  and measured its contribution to the energy density, ${E^\ell[\{\sigma_i \}] = -\frac{1}{N}\sum _{i<j}   J_{ij}^\ell \,\sigma_i\, \sigma_j \ .}$ This quantity is compared with the theoretical value ${-\frac 12 \int \mathrm{d}x\, \mathrm{d}y \; \rho(x) J_w(x-y) \rho(y)}$.

\begin{figure}
\begin{center}
\includegraphics[width=\linewidth]{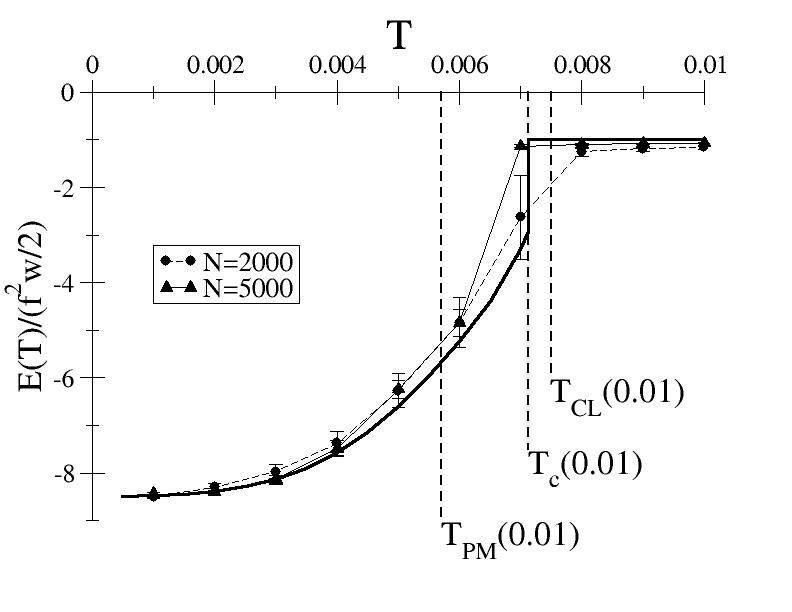}
\caption{Density of energy in the environment coherent with the clump for constant $\alpha=0.01$ (same realization of the disorder): results of Monte Carlo simulations for $N=2000$ (circles) and $N=5000$ (triangles) with error bars, compared to analytical result computed with $M=1000$ (line).}
\label{fig:EdeT}
\end{center}
\end{figure}

\begin{figure}
\begin{center}
\includegraphics[width=\linewidth]{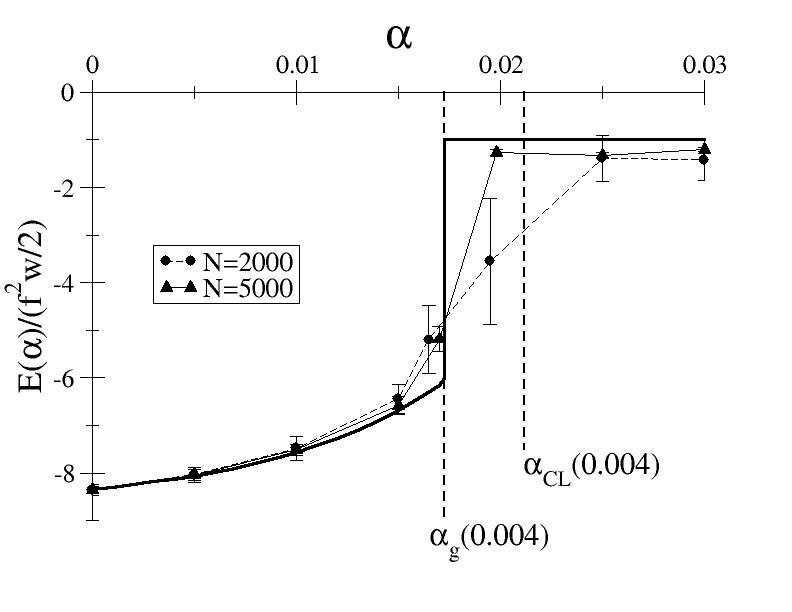}
\caption{Density of energy in the environment coherent with the clump for constant $T=0.004$: results of Monte Carlo simulations for $N=2000$ (circles) and $N=5000$ (triangles) with error bars, compared to analytical result computed with $M=1000$ (line).}
\label{fig:Edea}
\end{center}
\end{figure}

We have run simulations for different temperatures and numbers of environments, with $N=2000$ and $N=5000$ units. After thermalization, the energy of the coherent environment is recorded after $10^2$ rounds of $10N$ Monte Carlo steps each. Results are shown in Fig.~\ref{fig:EdeT} and Fig.~\ref{fig:Edea}.

\begin{figure}
\begin{center}
\includegraphics[width=\linewidth]{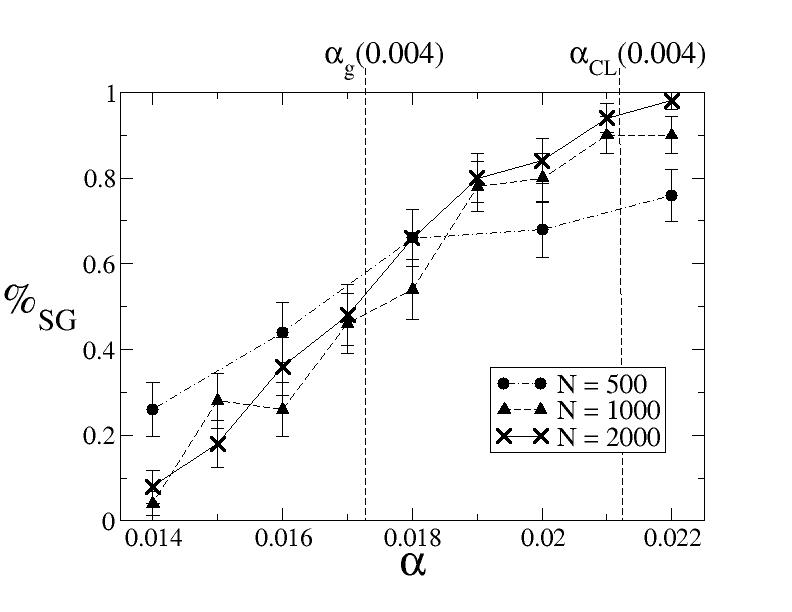}
\caption{Monte-Carlo simulations around the clump-glass transition for $T=0.004$: fraction of simulations found in the glassy phase after 100 rounds of $10N$ steps, as a function of $\alpha$ and for different $N$, with error bars. For each point the fraction was calculated from 50 simulations, half of which were started in a clump configuration and the other half in a uniform configuration.}
\label{fig:mctransi}
\end{center}
\end{figure}

The match with theoretical predictions is very good in the case of the clump-paramagnetic transition (Fig.~\ref{fig:EdeT}). Concerning the clump-glass transition (Fig.~\ref{fig:Edea}), as we mentioned above we expect the transition to occur at larger load, ${\alpha_g(T)<\alpha_g^{\text{observed}}<\alpha_{CL}(T)}$, due to the replica-symmetry broken nature of the glass phase. This expectation is corroborated by Fig.~\ref{fig:mctransi}, which represents the fraction of simulations ending in the glassy phase as a function of $\alpha$ for $T=0.004$. We have checked that this fraction does not depend on the initial conditions of the simulation. The transition occurs around ${\alpha\simeq0.018\pm0.001}$ (uncertainty due to long thermalization times in the simulations), while $\alpha_g \simeq 0.0173$ for $T=0.004$ used in the simulation.

\section{Choice of parameters}\label{parameters}

All the numerical computations above were performed with parameters values $w=0.05$ and $f=0.1$. The scope of the model is to account for the main qualitative properties observed in hippocampal recordings, so there is some arbitrariness in the choice of these values that we will discuss hereafter. To gain insight on the influence of the parameters on the behaviour of clump phase, we focus on two quantities representing its stability domain, namely $\alpha_{\text{CL}}$ and $T_{\text{CL}}$, respectively the load at which the clump phase becomes unstable at $T=0$ and the temperature at which the clump phase is unstable when $\alpha=0$. We also study the influence of $w$ and $f$ on first-order transitions, through $\alpha_g$ and $T_c$, respectively the load of transition to the glassy phase at $T=0$ and the temperature of transition to the PM phase at $\alpha=0$.

 \subsection{Size of place fields $w$}
\begin{figure}
   \centering
   \includegraphics[width=\linewidth]{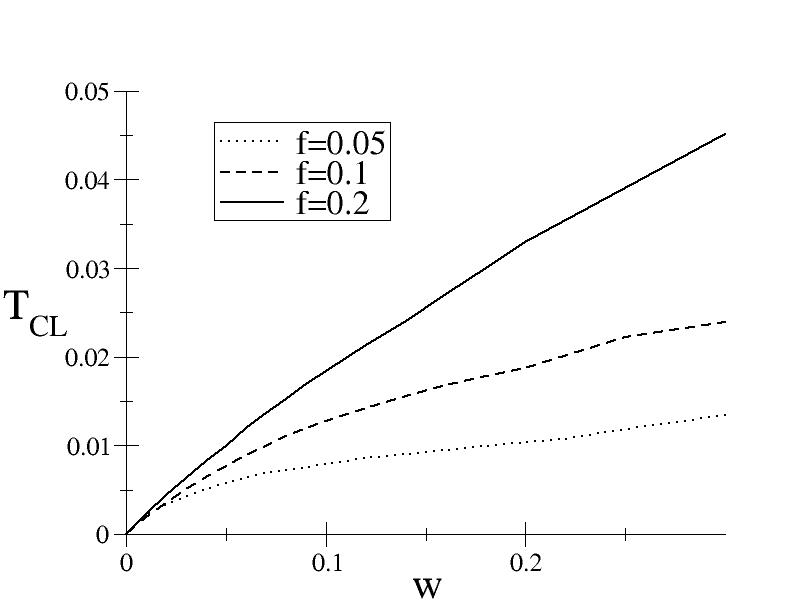}
   
   \includegraphics[width=\linewidth]{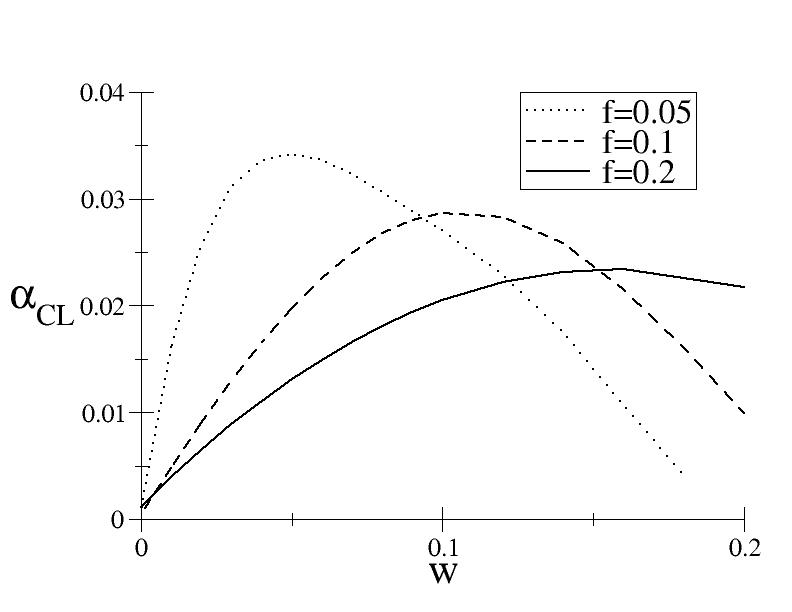}
 \caption{Influence of $w$ on the clump phase: $T_{CL}$ (top) and $\alpha_{CL}$  (bottom) as a function of $w$, for different fixed values of $f$.  Note the maximum around $w\sim f$ in the latter graph. Computations were done with $M=1000$. The numerical error is $\delta\alpha_{CL}\sim0.005$. }
\label{fig:Tdew}
 \end{figure}

Parameter $w$ is defined as the size of the place fields in relation to the size of the environment; hence $w$ defines the range of interactions resulting from the learning process. It fixes the width of the clump in the phase of localized activity. Experiments on rats have shown that the size of place fields depends on the size and complexity of the environment and on the behavioral context. A value $w=0.05$, \emph{i.e.} place fields occupying a few percents of the total space, is reasonable \cite{Amaral90}. We have varied $w$ for different values of $f$, and have found that $T_{\text{CL}}$ is a monotonously increasing function of $w$ (Fig.~\ref{fig:Tdew}). This result agrees with the intuition that increasing $w$ makes the clump phase more favorable energetically. It also appears that $\alpha_{\text{CL}}(w)$ has a maximum around $w\sim f$. In terms of storage capacity, this result suggests that there exists an optimal choice for the parameters: for a given level of inhibition hence a given number $fN$ of active neurons, choosing $w\sim f$ maximizes the proportion of these active neurons that are located in the place field. Given that the quenched noise coming from other environments is constant over space (see Fig.~\ref{fig:r}), $w\sim f$ is a trade-off between limiting the cross-talk and using the active neurons in the area covered by the place field.

\begin{figure}
   \centering
   \includegraphics[width=\linewidth]{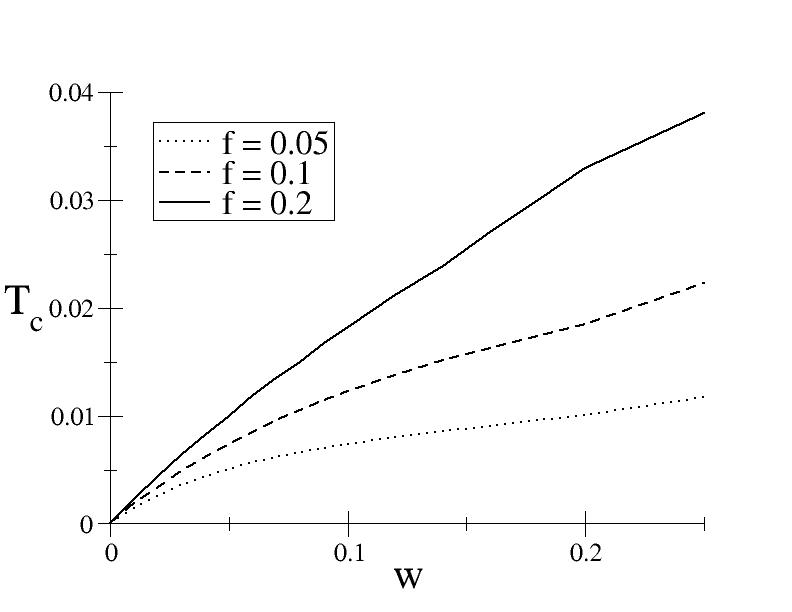}
   
   \includegraphics[width=\linewidth]{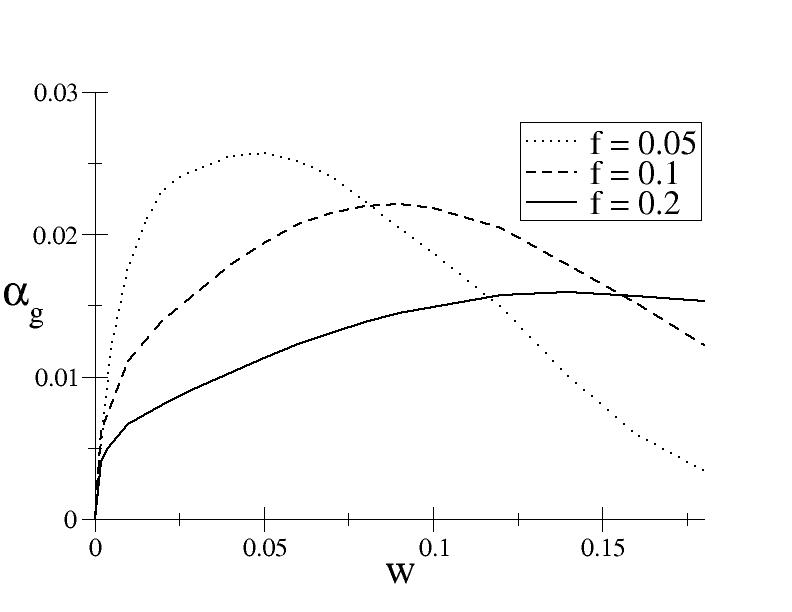}
 \caption{Influence of $w$ on the first-order transitions: $T_{c}$ (top) and $\alpha_{g}$  (bottom) as a function of $w$, for different fixed values of $f$. Computations were done with $M=1000$. }
\label{fig:Tcdew}
 \end{figure}

As far as thermodynamic transitions to the glassy and PM phases are concerned we find that $T_c$ and $\alpha_g$ behave similarly to, respectively, $T_{CL}$ and $\alpha_{CL}$ when $w$ varies, as shown in Fig.~\ref{fig:Tcdew}. Consequently, the qualitative aspect of the phase diagram remains the same when $w$ varies.

  \subsection{Total activity $f$}
\begin{figure}
 \centering
   \includegraphics[width=\linewidth]{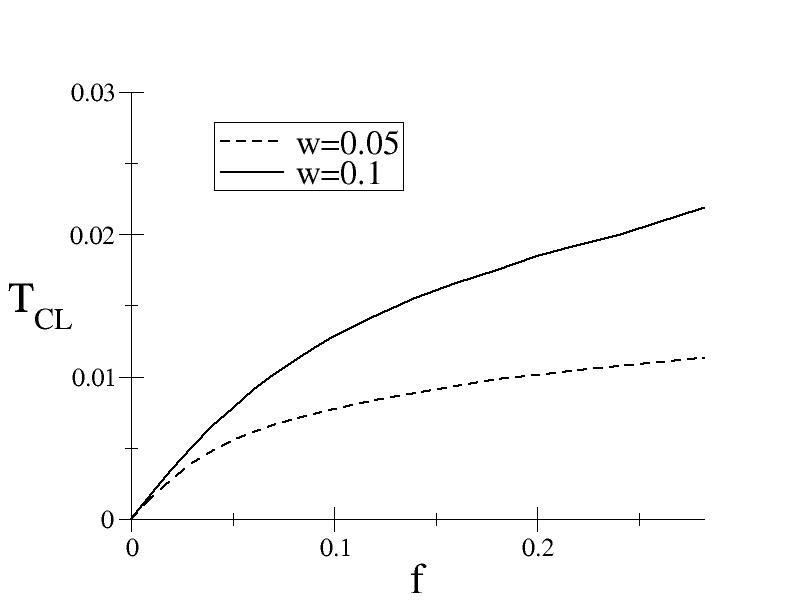}
   
   \includegraphics[width=\linewidth]{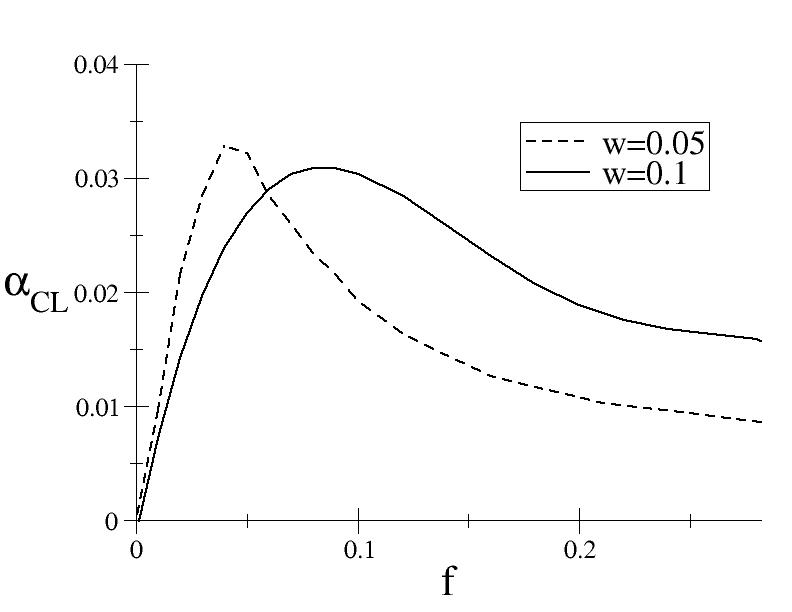}
 \caption{Influence of $f$ on the clump phase: $T_{CL}$ (top) and $\alpha_{CL}$  (bottom) as a function of $f$, for different fixed values of $w$. Note the maximum around $f\sim w$ in the latter graph. Computations were done with $M=1000$. The numerical error is $\delta\alpha_{CL}\sim0.005$.  }
\label{fig:Tdef}
 \end{figure}

\begin{figure}
 \centering
   \includegraphics[width=\linewidth]{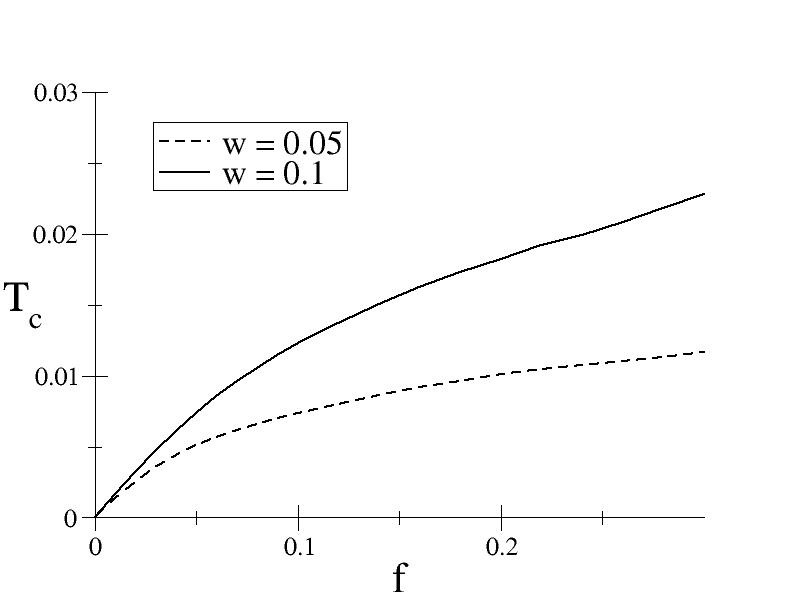}
   
   \includegraphics[width=\linewidth]{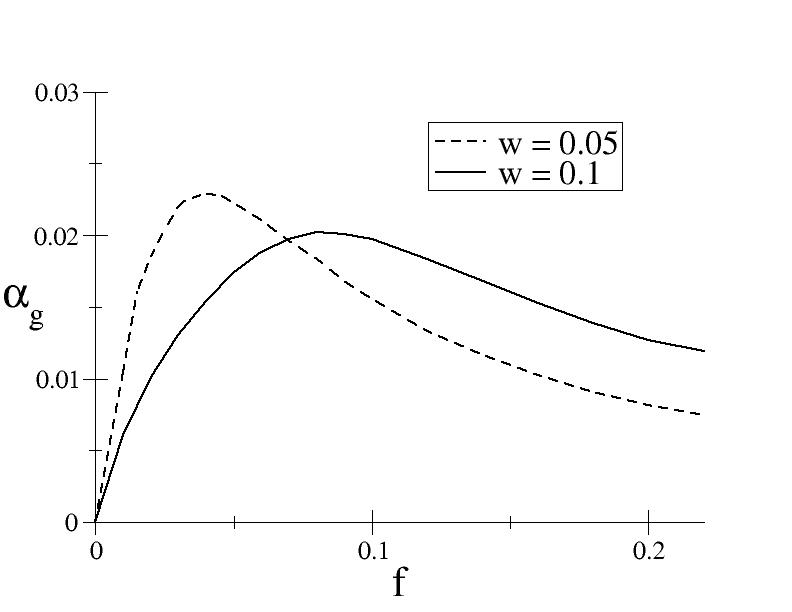}
 \caption{Influence of $f$ on the first-order transitions: $T_{c}$ (top) and $\alpha_{g}$  (bottom) as a function of $f$, for different fixed values of $w$.  Computations were done with $M=1000$. }
\label{fig:Tcdef}
 \end{figure}

Parameter $f$ is the activity level of the network fixed by global inhibition. As expected, $T_{\text{CL}}$ is a monotonously increasing function of $f$ (Fig.~\ref{fig:Tdef}). We find again a maximum of $\alpha_{\text{CL}}$ when $f$ is of the order of $w$, consistently with the previous results.  We also find that the boundary of the transition lines in phase diagram, $\alpha_g$ and $T_c$, behave similarly to $\alpha_{CL}$ and $T_{CL}$ (Fig.~\ref{fig:Tcdef}).

\section{Extensions and discussion}\label{discussion}

\subsection{Taking silent cells into account}\label{silentcells}
Thompson and Best \cite{ThompsonBest89} reported that not all pyramidal cells have place fields in a given environment: a significative fraction of them (63\% in their recording in CA1) being silent in this particular environment.
To take this effect into account, our model can be further refined to incorporate partial activity of the cell ensemble. We assume a fraction $1-c$ of silent cells, each environment being encoded by a subset of $cN$ units:
\begin{itemize}
\item In the reference environment (environment 0), $cN$ given spins $\sigma_i$ among the $N$ are assigned regurlarly spaced place field centers $p(i)$ and they interact through the coupling matrix 
\begin{equation}
J^0_{kl} = \left\{\begin{array} {c c }
\frac {1}{N} & \hbox{\rm if}\  d(k,l) \le \frac {wN}{2}\\
0 & \text{otherwise} \end{array} \right. \ .
\end{equation}
The contribution to the energy is
\begin{equation}
E^0[\{\sigma_i \}] = -\sum _{i<j\le cN}  J_{p(i)p(j)}^0 \,\sigma_i\, \sigma_j \ .
\end{equation} 
\item In each of the $L$ other environments, each spin $\sigma_i$ (of the all $N$ spins) is selected with probability $c$ and the place field centers are reshuffled by random permutation $\pi^{\ell}$. For each $i$ let the random dilution variable
\begin{equation}
\tau_i^{\ell} = \left\{\begin{array} {c r }
1 & \hbox{\rm with probability}\  c\\
0 & \hbox{\rm with probability}\  1-c  \end{array} \right. \ .
\end{equation} 
The corresponding energy writes
\begin{equation}
E ^{\ell}[\{\sigma_i \}] = -\sum _{i<j}  J_{ij}^0 \,\tau_{\pi^{\ell}(i)}^{\ell}\sigma_{\pi^{\ell}(i)}\, \tau_{\pi^{\ell}(j)}^{\ell}\sigma_{\pi^{\ell}(j)} \ .
\end{equation} 
\end{itemize}
We incorporate this new hypothesis in the calculation of the average over disorder of the replicated partition function. The average is now over two types of disorder: the permutations $\pi^{\ell}$ and the selection of involved cells $\tau_i^{\ell}$.

 We still consider configurations $\{\sigma_i^a\}$ satisfying condition (\ref{activity}).  Moreover, for each realization of the $\tau_i$ we restrict the sum over configurations satisfying
\begin{equation}
\frac{1}{cN}\sum_{i=1}^{N}\tau^{\ell}_i\sigma^a_{i}=f
\end{equation}
that is, the global inhibition is homogeneously distributed over the different subpopulations of neurons. This hypothesis  is reasonable regarding the effective action of inhibitory cells on the network that we want to model. We can show that, at the order $\frac{1}{N^2}$ it is always true if (\ref{activity}) is satisfied. 

\begin{figure}
 \centering
   \includegraphics[width=\linewidth]{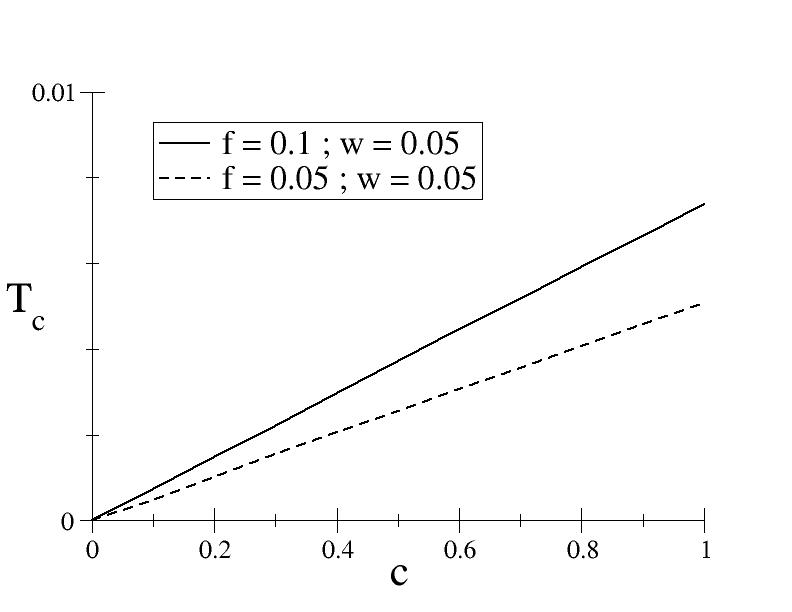}
   
   \includegraphics[width=\linewidth]{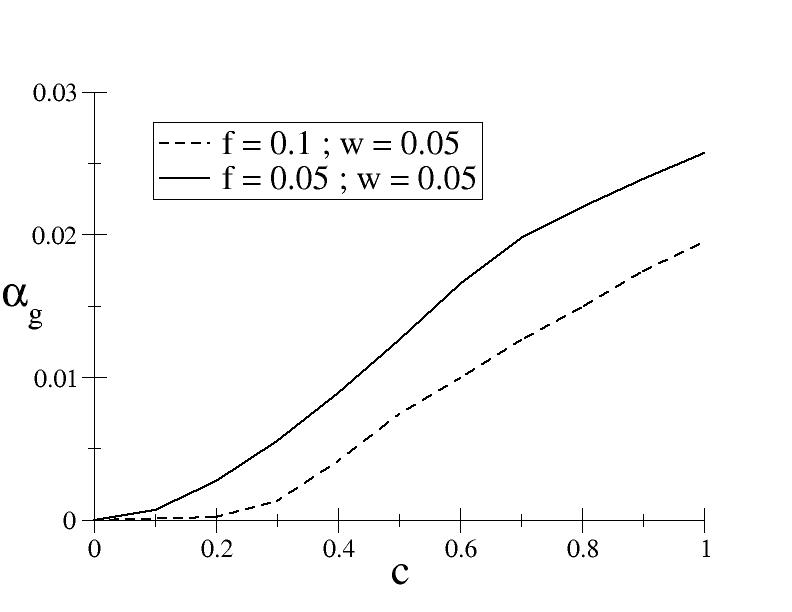}
 \caption{Effect of partial activity: Influence of the fraction $c$ of active cells on the clump domain: $T_c$ (top) and $\alpha_g$  (bottom) as a function of $c$, for different fixed values of $f$ and $w$. Computations were done with $M=200$.   }
\label{fig:effetc}
 \end{figure}

The calculation, detailed in appendix \ref{appact}, follows the same steps as in the $c=1$ case. The only difference is that one has to be careful when averaging over the two different disorders: we first perform the average over permutations for a given realization of the $\tau_i^{\ell}$, and then over the dilution variables $\tau_i^{\ell}$. We obtain the following expression for the density of free energy in one dimension:
\begin{align}\label{eq:Fc}
{\cal F}_c &= \frac {\alpha \beta}2 r(f-q) -\frac{\alpha}{\beta}\psi_c(q,\beta)+c\int \mathrm{d}x\, \mu(x)\rho(x)+(1-c)\mu_2f\nonumber\\
&-\frac {c^2}{2} \int \mathrm{d}x\, \mathrm{d}y\, \rho(x)J_w(x-y)\rho(y)-\lambda c\big(\int\mathrm{d} x\, \rho(x)-f\big)\nonumber\\
&-\frac c\beta\int \mathrm{d}x \int Dz \log \bigg( 1 + e^{\beta
z\sqrt{\alpha r} + \beta\mu(x)}\bigg)\nonumber\\
&-\frac {(1-c)}{\beta} \int Dz \log \bigg( 1 + e^{\beta
z\sqrt{\alpha r}+\beta\mu_2 }\bigg)
\end{align}
where $q$ is defined as before and 
\begin{align}
 \psi_c(q,\beta)& =\sum\limits_{k\geq 1}\left[\frac{\beta c(q-f^2)\sin(k\pi w)}{k\pi- \beta c(f-q)\sin(k\pi w)}\right. \nonumber \\
&-\left.\log\left(1-\frac{\beta c(f-q)\sin(k\pi w)}{k\pi}\right)\right]\ .
\end{align}
The optimization equations are 
\begin{align}
\label{eq:saddlepoint2}
&r=2c^2(q-f^2)\sum\limits_{k\geq 1}\left[\frac{k \pi}{\sin(k\pi w)}-\beta c(f-q)\right]^{-2}\ ,\ \nonumber\\
&q=c\int\mathrm{d}x\int\mathrm{D}z[1+e^{-\beta z\sqrt{\alpha r}-\beta\mu(x)}]^{-2}\nonumber\\
&\ \ +(1-c)\int\mathrm{D}z[1+e^{-\beta z\sqrt{\alpha r}-\beta\mu_2}]^{-2},\nonumber\\
&\rho(x)=\int\mathrm{D}z[1+e^{-\beta z\sqrt{\alpha r}-\beta\mu(x)}]^{-1},\nonumber\\
&\mu(x)=c\int\mathrm{d}y\, J_w(x-y)\rho(y)+\lambda\ ,\nonumber\\
&\int\mathrm{d}x\, \rho(x)=f\ ,\nonumber\\
&f=\int\mathrm{D}z[1+e^{-\beta z\sqrt{\alpha r}-\beta\mu_2}]^{-1}\ .
 \end{align}
In the partial activity model, the active spins (with activity $\rho(x)$) obey equations that are very similar to the previous case, with a dilution factor coming from the silent spins which are in a paramagnetic phase. From a qualitative point of view the behavior of the system does not differ significantly from the system with all spins active ($c=1$). We have computed the effect of varying $c$ on the value of $T_c$ and $\alpha_g$: $T_c$ is found to be a linear function of $c$, while  $\alpha_g$ is a monotonously increasing function of $c$. Results are shown in Fig.~\ref{fig:effetc}. 

\subsection{Relationship with linear threshold models and previous studies}
Several attractor neural network models for hippocampus have been proposed in previous works. Tsodyks \& Sejnowski \cite{TsodyksSejnowski95} proposed a rate model with semi-linear threshold neurons, uniform inhibition and excitatory synapses between neurons with neighbouring place fields, with a strength decaying exponentially  with distance. Their study was limited to the single environment, one-dimensional case. They showed the formation of localized activity. Moreover, they studied the effect of inhomogeneities in the synaptic matrix due to irregularities in the learning process, an interesting effect that we do not address here.

Battaglia \& Treves \cite{BattagliaTreves98} introduced the multiple environment storage in additive synapses. They studied the case of linear threshold neurons with generic form of kernel of connection weights. The free-energy is calculated  implementing the threshold linear transfer function and averaging over disorder in the replica-symmetric approximation, along the lines developed in \cite{Treves90}. The clump phase is studied at zero temperature, and the storage capacity is found as the maximal value of $\alpha$ for which localized solutions exist. Different forms of couplings and varying sparsity of the representation are considered, and a enlightening parallel with episodic memory is proposed. The issue of information storage is addressed.

 Our method is in the same spirit as \cite{BattagliaTreves98}, but the model differs as we consider binary units instead of threshold linear units (i.e. without saturation) for a simple coupling matrix and an explicit form of inhibition. Nevertheless, a parallel can be drawn between the range of interaction $w$ in our model and the 'map sparsity' $\frac{1}{|M|}$ in  \cite{BattagliaTreves98}. In spite of the differences between the models, the order of magnitude of the maximal storage capacity is the same in both models: $\sim 3.10^{-2}$ in 1D, $\sim 8.10^{-3}$ in 2D (see Fig.~1 and 2 in \cite{BattagliaTreves98}). The 'chart sparsity' $\alpha_c$ in  \cite{BattagliaTreves98} corresponds to our parameter $c$.

The main difference between both models lays in the way noise is taken into account. In \cite{BattagliaTreves98}, the level of noise is embedded in the rate model, in the gain $g$ of the units, and is not taken into account in the thermodynamics since the study is carried out at zero temperature. Our model considers binary units with a level of noise $T$ corresponding to the thermodynamic temperature. On average binary neurons behave as rate neurons with sigmoidal transfer function of gain $\frac1T$ (see section \ref{linkwithrate}). From this point of view our model is more microscopic than the one in \cite{BattagliaTreves98}, as we have a description of noise at the neuron level. Furthermore, we have looked at the stability of the clump phase against replicon modes. Our study also includes the other regimes of activity of the model (i.e. the PM and SG phases) and their thermodynamic stability compared to the clump phase, summarized in the phase diagram.

\subsection{Conclusion}

In this paper we have introduced an attractor neural network model for the storage of multiple spatial maps in the hippocampus. Although very simplified, the model accounts for experimentally observed properties of place cells, such as the remapping of place fields from one environment to the other. We showed that multiple maps can be simultaneously learned in the same network, {\em i.e.} with the same synaptic coupling coefficients, even in the presence of noise in the neural response. Remarkably, moderate levels of noise can even slightly increase the capacity storage with respect to the noiseless case. Notice that the qualitative behaviour of the model is robust to changes in the value of the parameters; for instance we do not expect that changing the couplings from a square-box function into an exponentially decreasing function over the distance $w N$ in $D=1$ or $\sqrt{w N}$ in $D=2$ would affect much the phase diagram. 

The storage of a map manifests itself through the fact that the neural activity is localized, and acquires a clump-like shape in the corresponding environment. When the load (number of environments) or the noise are too high the neural activity cannot be localized any longer in any one of the environments. For high noise, the activity, averaged over time, simply becomes uniform over the space. For high loads the activity is not uniform, but is delocalized with spatial heterogeneities controlled by the cross-talks between the (too many) maps. The prevalence of the glassy phase at high load and of the uniform (paramagnetic in the physics language) phase at high noise moderately limits the extension of the clump phase. Moreover, we have found that in the glassy phase the replica symmetric assumption is not correct, and we may expect from general consideration about replica symmetry-breaking that the first-order transition from the clump phase to the glassy phase occurs at higher loads $\alpha$. Remarkably the clump phase is therefore the thermodynamically dominant phase in nearly all of its stability domain.

Our work would deserve to be extended along other directions. First the assumption that synaptic couplings additively sum up the contributions coming from all the environments could be lifted. We could replace the synapses $J_{ij}$ with non-linear function $G(J_{ij})$. The additive case corresponds to $G(x)=x$, while  a strongly non-additive synapse is obtained with the choice ${G(x)=\min (x,\frac 1N)}$: synapses can be written in only once, and contributions from different environments do not add up but saturate the synaptic coupling. It would be worth extending the study of nonlinear synapses done for the Hopfield model \cite{Sompolinsky86,VanHemmen87} to the present model. 

Secondly we have considered that the only source of (quenched) noise was the interference between the multiple environments. In other words, in the single-environment case, our synaptic matrix is translationally invariant and the center of the activity clump can be moved at no energy cost in space. This idealizing assumption was done to study the effect of multiple-environment cross-talk only. However, even in the single environment case, place fields do not define a perfectly regular covering of space. We expect that such heterogeneities in the couplings will further destabilize the clump phase, and decrease the storage capacity \cite{Sompolinsky86}. Quantifying those effects would be interesting. 

However the most important extension seems to us to be the study of the dynamics. The richness of the phase diagram we have unveiled here and the multiplicity of phases for the system raise the question of if and how the network activity makes transitions between those phases. Multiple environments stored in the same network not only influence the shape of the clump and lead to transitions to a glassy phase, but they can as well provoke transitions between attractors. The study of these transitions, of the corresponding reaction paths will be reported in a companion paper \cite{Monasson13}. It could reveal useful to interpret recent experiments, where changes of the hippocampal activity resulting from the 'teleportation' of the rat have been recorded \cite{Jezek11}. In addition it would be interesting to understand in a more quantitative way the activated diffusion process of the clump in an environment. Under the presence of other maps, the clump does not freely diffuse and quantifying the barriers opposing motion, as 
well as understanding the qualitative difference between motions in 1D and 2D spaces would be very useful. 

\section*{Acknowledgements}

We are deeply indebted to J. Hopfield for enlightening discussions, without which this work would not have been possible. We are grateful to N. Brunel for a critical reading of the manuscript and to S. Cocco, F. Stella, A. Treves for very useful discussions. R.M. acknowledges the hospitality and the financial support of the Simons Center for Systems Biology, Institute for Advanced Study, Princeton, where an initial part of this work was done. The work of S.R. is supported by a grant from D\'{e}l\'{e}gation G\'{e}n\'{e}rale de l'Armement.

\appendix

\section{Formulas for two-dimensional maps}

\subsection{Single environment - Stability of the PM phase}
\label{app2D1env}

With periodic boundary conditions we can write the Fourier expansion 
\begin{equation}
 \delta\rho(\vec{x})=\sum\limits_{\vec{k}}\delta\hat{\rho}(\vec{k})e^{i\vec{k}\cdot\vec{x}}
\end{equation}
where the sum runs over vectors $\vec{k}$ belonging to the reciprocal lattice and $\vec{k}\neq\vec{0}$ because of constraint (\ref{constraint1}).\\
To simplify the computation we replace the disk of interaction by a square: $J_w(\vec{u})=1$ if $|u_x|$ and $|u_y|<\frac {\sqrt{w}}2$, so that 
\begin{eqnarray*}
 \sum\limits_{\vec{k}}\int\limits_{|\vec{x}-\vec{y}|<\frac {\sqrt{w}}2}\mathrm{d}\vec{y}\,e^{i\vec{k}\cdot\vec{y}}=\sum\limits_{\underset{\neq(0,0)}{k_1,k_2}}e^{i\vec{k}\cdot\vec{x}}\frac{\sin(k_1\pi\sqrt{w})\sin(k_2\pi\sqrt{w})}{k_1k_2\pi^2}
\end{eqnarray*}
Therefore all eigenvalues are positive provided that ${T>T_{PM}^{\text{2D}}}$ where
\begin{align}
 T_{PM}^{\text{2D}}&=f(1-f)\max_{\underset{\neq(0,0)\}}{\{k_1,k_2}}\bigg(\frac{\sin(k_1\pi\sqrt{w})\sin(k_2\pi\sqrt{w})}{k_1k_2\pi^2}\bigg)\nonumber\\
&=f(1-f)\sqrt{w}\frac{\sin(\pi\sqrt{w})}{\pi}\ .
\end{align}
In the case of $w<<1$, $T_{PM}^{\text{2D}}\approx T_{PM}^{\text{1D}}$.

\subsection{Order parameters for multiple environments}\label{app2D}

The only difference in the replica computation lays in the eigenvalues of the coupling matrix. Thus, in dimension 2, the free energy functional writes:
\begin{eqnarray}\label{calf2d}
{\cal F}^{\text{2D}} &=& \frac {\alpha \beta}2 r(f-q) -\frac\alpha\beta\psi^{\text{2D}}(q,\beta) \nonumber\\
&-&\frac 12 \int \mathrm{d}\vec{x}\, \mathrm{d}\vec{y}\, \rho(\vec{x})J_w(\vec{x}-\vec{y})\rho(\vec{y}) +\int \mathrm{d}\vec{x}\, \mu(\vec{x})\rho(\vec{x})\nonumber  \\
&-&\frac 1\beta\int \mathrm{d}\vec{x}\, \int Dz \log ( 1 + e^{\beta
z\sqrt{\alpha r} + \beta\mu(\vec{x})})\ ,
\end{eqnarray}
where
\begin{align}
\psi^{\text{2D}}(q,\beta)\equiv2\sum_{\underset{\neq(0,0)}{(k_1,k_2)}}&\bigg(\frac{\beta(q-f^2)}{\phi(k_1,k_2)-\beta(f-q)}\nonumber\\
&-\log(1-\frac{\beta(f-q)}{\phi(k_1,k_2)})\bigg)
\end{align}
with

\begin{align}
 \phi(k_1,k_2)\equiv\frac{k_1k_2\pi^2}{\sin(k_1\pi \sqrt{w})\sin(k_2\pi \sqrt{w})}
\end{align}

Hence the saddle point equations write:
\begin{align}
\label{eq:saddlepoint2d}
&r=4(q-f^2)\sum\limits_{(k_1,k_2)\neq(0,0)}\left(\phi(k_1,k_2)-\beta(f-q)\right)^{-2},\nonumber\\
&q=\int\mathrm{d}\vec{x}\int\mathrm{D}z[1+e^{-\beta z\sqrt{\alpha r}-\beta\mu(\vec{x})}]^{-2},\nonumber\\
&\rho(\vec{x})=\int\mathrm{D}z[1+e^{-\beta z\sqrt{\alpha r}-\beta\mu(\vec{x})}]^{-1},\nonumber\\
&\mu(\vec{x})=\int\mathrm{d}\vec{y}\,J_w(\vec{x}-\vec{y})\rho(\vec{y})+\lambda\ .
    \end{align}
   where $\lambda$ is determined to enforce constraint (\ref{constraint1}).

    In the $D=2$ case equations (\ref{extr}) can be simplified by exploiting the invariance by rotation: in polar coordinates 
\begin{align}
 \mu^{2D}(r)&=2\int_{\underset{|r-r'|\leq\sqrt{\frac {w}{\pi}}}{r+r'\geq\sqrt{\frac {w}{\pi}}}}\mathrm{d}r'\rho^{2D}(r')r'\arccos\left(\frac{r^2+r'^2-\frac{w}{\pi}}{2rr'}\right)\nonumber\\
& +2\pi\int\limits_{ r+r'<\sqrt{\frac {w}{\pi}}}\mathrm{d}r'\rho^{2D}(r')r'+\lambda 
\end{align}

\begin{figure}
   \centering
   \includegraphics[width=\linewidth]{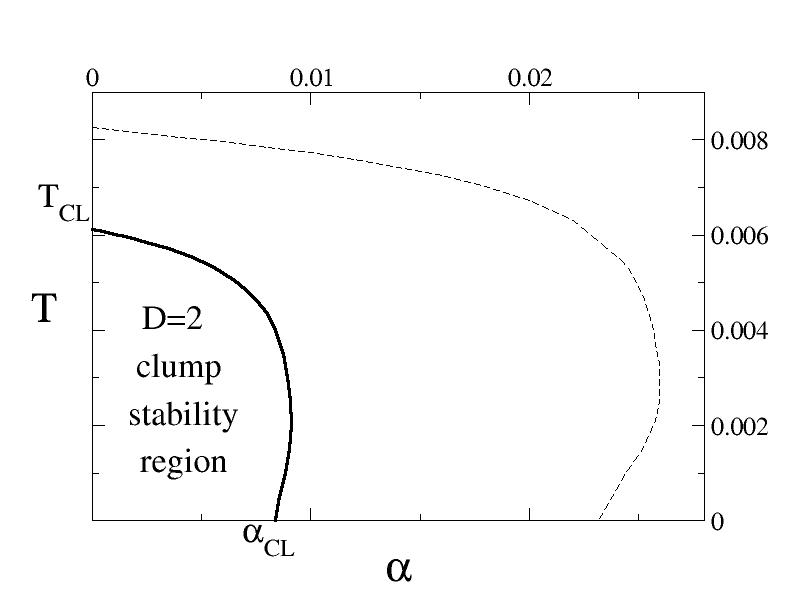}
   \label{fig:stab2d}
 \caption{Solid line: Longitudinal stability region of the clump phase for ${D=2}$. The ${D=1}$ case is shown in thin dashed line for comparison.}
 \end{figure}

We thus computed $\rho(r)$ in the clump phase and found the region in the $(\alpha,T)$ plane where this solution is stable against longitudinal modes. We find that this region is reduced compared to the $D=1$ case, but its shape is qualitatively similar. The result is shown in figure (\ref{fig:stab2d}).

\section{Average of the Boltzmann factor over a random environment}
\label{appaverage}

The purpose of this appendix is to calculate
\begin{eqnarray}
\Xi \big(\vec{\boldsymbol \sigma}\big) &=& \frac 1{N!} \sum _{\pi} \exp\left[\beta \sum_{i<j} J^0_{ij} \sum_{a=1}^n \sigma_{\pi(i)}^a \sigma_{\pi(j)} ^a\right]\nonumber \\
&=&C \xi \big(\vec{\boldsymbol \sigma}\big)
\end{eqnarray}
with
\begin{align*}
C \equiv\exp \left(-\frac \beta 2 n f(1-f) +N\frac \beta 2 n w f^2\right) \ , 
\end{align*}
\begin{align}
\xi (\vec{\boldsymbol \sigma}) \equiv \frac 1{N!} \sum _{\pi} \exp\big[\frac \beta 2 \sum_{i,j} J^0_{ij} \sum_{a=1}^n (\sigma_{\pi(i)}^a -f)(\sigma_{\pi(j)} ^a-f)\big] 
\end{align}
where the sum is carried out over all permutations of $N$ elements.\\
The eigenvectors of the matrix $J^0$ are plane waves. Let $v_{q,j}$ denote the $j^{th}$ (real-valued) component of the $q^{th}$ normalized eigenvector, and $\lambda_q$ the associated eigenvalue. Then,
\begin{equation}\label{inter1}
\sum_{i,j} J^0_{ij}  (\sigma_{\pi(i)}^a -f)(\sigma_{\pi(j)} ^a-f) = \sum _{q=1}^{N-1} \lambda_q \bigg( \sum_j v_{q,j}\, ( \sigma_{\pi(j)}^a -f) \bigg) ^2 \ .
\end{equation}
Due to condition (\ref{activity}) we have discarded the homogeneous mode $q=0$ from the sum in (\ref{inter1}). Introducing a set of  $n(N-1)$ independent Gaussian variables with zero mean and variance unity, denoted by $\Phi_q^a$, we can write (all odd powers of $\sqrt{\beta}$ vanish after integration over the Gaussian measure)
\begin{align}\label{xi0}
\xi (\vec{\boldsymbol \sigma}) &=& \left\langle \exp\left[\sqrt \beta \sum\limits_{q,a,j} \sqrt{\lambda_q}\,v_{q,j}\, \Phi^a_{q}\,(\sigma_{\pi(j)}^a -f)\right]\right\rangle_{\pi, \Phi}\nonumber \\
&=&1+ \sum\limits _{k\ge 1} \frac{\beta^k}{(2k)!} \sum_{\underset{i=1\cdots2k}{q_i,a_i,j_i}} \Big( v_{q_1,j_1}\, v_{q_2,j_2}\ldots v_{q_{2k},j_{2k}}\nonumber \\ 
&\times&\sqrt{\lambda_{q_1}\lambda_{q_2}\ldots \lambda_{q_{2k}}}\ T_{j_1,j_2\ldots j_{2k}}^{a_1,a_2\ldots a_{2k}}\langle  \Phi^{a_1}_{q_1}\,\Phi^{a_2}_{q_2}\ldots \Phi^{a_{2k}}_{q_{2k}}\rangle_\Phi\Big) 
\end{align}
where 
\begin{equation}\label{defTfactor}
T_{i_1,i_2,\ldots, i_{2k}}^{a_1,a_2,\ldots, a_{2k}} \equiv \langle  (\sigma_{\pi(i_1)}^{a_1} -f)(\sigma_{\pi(i_2)}^{a_2} -f)\ldots (\sigma_{\pi(i_{2k})}^{a_{2k}} -f)\rangle _\pi \ . 
\end{equation}
Using Wick's theorem the $2k$--point correlation function of the $\Phi$ variables is easy to calculate. The outcome is a multiplicative factor $(2k-1)!!$, and the replacement of the $2k$ sums over the indices $q_m,a_m$ by only $k$ independent sums over $q_m,a_m$.  The value of $T$ (\ref{defTfactor}) depends only on the number $M$ of distinct indices, $i_m$, and of their associated multiplicities. 
Power counting shows that $T_{i_1,i_2,\ldots, i_{2k}} ^{a_1,a_2,\ldots, a_{2k}} $ vanishes in the infinite $N$ limit unless the set $\{i_1,i_2,\ldots, i_{2k}\}$ includes exactly $k$ distinct indices, each one with multiplicity two. When this condition holds we write $(a_m,a'_m)$ the replica indices attached to the $m^{th}$ distinct index $i$, with $m=1,2,\ldots, k$. Then, in the large $N$ limit, 
\begin{equation}
T_{i_1,i_1,i_2,i_2\ldots, i_k,i_k}^{a_1,a'_1,a_2,a'_2,\ldots, a_k,a'_k} = \prod_{m=1}^k
\big( q^{a_m a'_m}-f^2\big)\ . 
\end{equation}
Nevertheless, when summing over $k$ in (\ref{xi0}), the inversion of limits $N\rightarrow\infty$ and $\sum\limits_{k=1}^\infty$ is allowed only in the cases where the $\sqrt{\beta\lambda_q} v_{q,j}(\sigma_{\pi(j)}^a-f)$ are not too large so that the decomposition converges quickly. This is not the case if by any chance the configuration $\vec{\boldsymbol \sigma}$ is coherent with the environment 0, because then a large number (of order $N$) of terms will have to be taken into account in the sum. The probability of such realizations is exponentially small in $N$. Since we are only interested in the environment 0 we can discard those terms and consider only the realizations for which the approximation above is correct.\\
We are left with the summation over the $j_m$ indices. Using the identities
\begin{equation}
\sum _j v_{q,j}\, v_{q',j} = \delta_{q,q'}\ ,
\end{equation}
 we obtain from (\ref{xi0}) the following expression
\begin{equation}\label{xi2}
\xi (\vec{\boldsymbol \sigma}) = 1+\sum _{k\ge 1}\ \frac{(\beta /2)^k}{k!}\sum \limits_{{\cal P} } w({\cal P}) \ ,
\end{equation}
where the last sum runs over all weighted pairings among $2k$ points, described as follows:
\begin{itemize}
\item we define $2k$ points. The first $k$ points carry the pair-indices $(q_m,a_m)$, with $m$ running from 1 to $k$. The second $k$ points carry the same pair-indices. Hence, each pair index $(q_m,a_m)$ is shared by two points.
\item a pairing ${\cal P}$ is a set of $k$ bonds ${b_\ell\equiv\{ (q_{m_\ell},a_{m_\ell}),(q_{m'_\ell},a_{m'_\ell})\}}$, $\ell=1,2,\ldots, k$, each joining one pair of points (dimer coverage).
\item the weight of the pairing is
\begin{equation}\label{weight}
w({\cal P}) \equiv \sum _{\underset{q_1,\cdots,q_{k}}{a_1,\cdots,a_{k}}} \;\prod_{m =1}^k \lambda_{q_m}\; \prod_{\ell =1}^k \delta _{q_{m_\ell},q_{m'_\ell}}\;  \big( q^{a_{m_\ell} a_{m'_\ell}}-f^2\big)\ .
\end{equation}
\end{itemize}
We denote ${\bf q}$ the overlap matrix with entries $q^{ab}$ and {\bf 1}$_n$ the $n\times n$ matrix whose all entries are equal to one.\\
Let's introduce a notation for the moments of the eigenvalues:
 \begin{equation}
\label{eqlambda}
 \Lambda_h \equiv \sum_{q\ge 1} \lambda_q ^h = 2 \sum_{q\ge 1} \left( \frac{\sin (q \pi w)}{q\pi}\right)^h \ .
 \end{equation}

\begin{figure}
  \includegraphics[width=.45\linewidth]{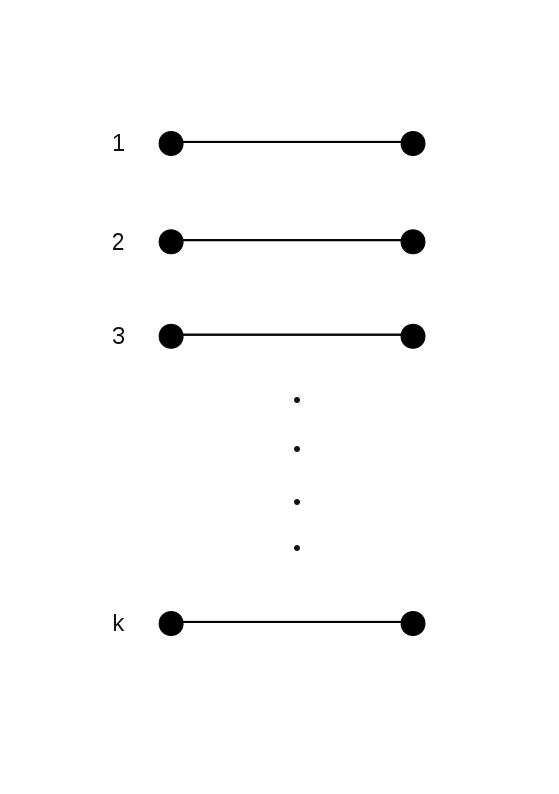}
  \includegraphics[width=.45\linewidth]{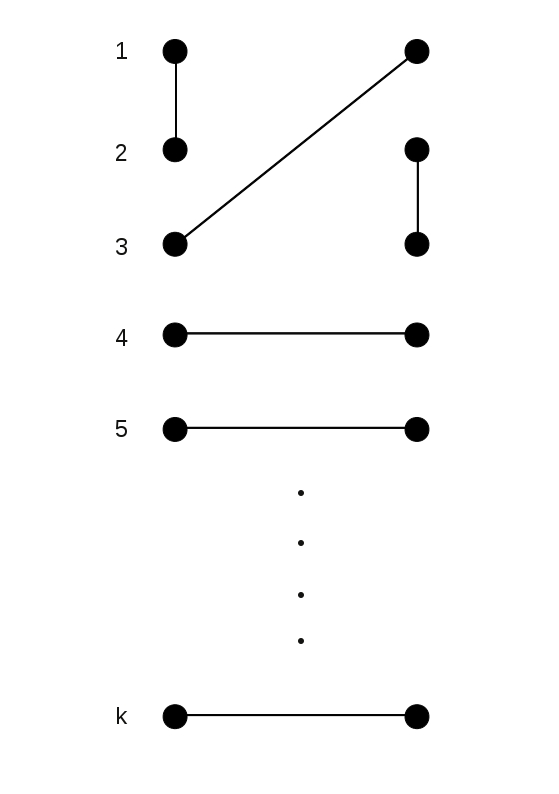}
  \caption{Two examples of pairings between $2k$ points: ${\cal P}_A$ (left) and ${\cal P}_B$ (right).}
\label{fig:examplepairings}
\end{figure}

Two examples of pairings are shown in Fig.~\ref{fig:examplepairings}. The  weight associated to the pairing ${\cal P}_A$ is simply
\begin{eqnarray}
w({\cal P}_A) &=& \prod_{m =1}^k \left[ \sum_{q_m} \lambda_{q_m}\sum _{a_m} \big( q^{a_m a_m} -f^2 \big) \right]\nonumber \\ &=&\big( \Lambda _1\, \hbox{Trace} [{\bf q} -f^2]\big)^k\nonumber\\
&=&  \big(\Lambda _1\, n\, f(1-f)\big)^k \ ,
\end{eqnarray}
as all Kronecker $\delta$ in (\ref{weight}) are equal to 1 by construction.
The weight associated to the second pairing in Fig.~\ref{fig:examplepairings} is
\begin{align}
w({\cal P}_B) 
=  \big(\Lambda _3\, \hbox{Trace} [({\bf q} -f^2)^3]\big)\,\big( \Lambda _1\, \hbox{Trace} [{\bf q} -f^2]\big)^{k-3}
\end{align}

For a given pairing, 
\begin{itemize}
 \item the horizontal bonds represent independent replicas : point number $m$ leads to a factor $\sum_{q_m} \lambda_{q_m}\sum _{a_m} \big( q^{a_m a_m} -f^2 \big)$ in the weight of the pairing.
\item the vertical and diagonal bonds couple replicas together
\end{itemize}

We then have to calculate the combinatorial multiplicity of the weights, \emph{i.e.} how many pairings have the same weight in the sum (\ref{xi2}). For a given $k$, a pairing associates points by groups of $j$ coupled replicas indices (i.e $2j$ points). Let $m_j$ be the number of such groups in a given pairing. We have $\sum_j jm_j=k$. Pairings $\cal P$ with the same $(j,m_j)$ have equal weights 
\begin{equation}
 w({\cal P})=w(\{(j,m_j)\})=\prod_j\left(\Lambda_j T_j \right)^{m_j}
\end{equation}
 where we set $T_j\equiv\hbox{Trace} [ ({\bf q} -f^2{\bf 1})^j]$.

 Combinatorial study shows that the number of such pairings is
\begin{equation}
 \mathcal{N}(\{(j,m_j)\})=k!\prod_j\frac{1}{m_j!}\left(\frac{2^{j-1}}{j}\right)^{m_j}
\end{equation}
Finally, using $\sum_j jm_j=k$ and (\ref{eqlambda}), we can rewrite
\begin{align}
\xi (\vec{\boldsymbol \sigma}) &= 1+\sum _{k\ge 1}\ (\frac{\beta}{2})^k\prod_j\sum_{m_j\ge 0}\frac{1}{m_j!}\left(\frac{2^{j-1}}{j}\right)^{m_j}(\Lambda_j T_j)^{m_j}\nonumber\\
&=\exp\Big[\sum_j\frac12\frac{\beta^j}{j}\Lambda_j T_j\Big]\nonumber\\
&=\exp\Big[-\sum_{\lambda \ne 0} \text{Trace} \;\log[\text{\bf Id}_n - \beta \lambda \big({\bf q}-f^2\, {\bf 1}_n\big)]\Big]
\end{align}

\section{Replica symmetric calculation of the free energy}
\label{appreplica}

We introduce parameters $r^{ab}$ conjugated to the overlaps $q^{ab}$. With this notation, we have (up to a multiplicative constant):
\begin{equation}
 \overline{Z^n} = \sum_{\vec{\boldsymbol \sigma}}\int\prod_{a<b}\mathrm{d} q^{ab}\mathrm{d} r^{ab} \exp\,[G(\{q^{ab},r^{ab}\},\vec{\boldsymbol \sigma})]
\end{equation}
where 
\begin{align}
 G(\{q^{ab},r^{ab}\},\vec{\boldsymbol \sigma})&=N\alpha\beta^2\sum\limits_{a<b}r^{ab}(\frac{1}{N}\sum\limits_i\sigma_i^a\sigma_i^b-q^{ab})\nonumber\\
&+\frac{\beta}{2N}\sum\limits_a\sum\limits_{|i-j|<\frac{wN}{2}}\sigma_i^a\sigma_j^a\nonumber\\
&-\alpha N\sum_{\lambda \ne 0} \text{Trace} \;\log[\text{\bf Id}_n - \beta \lambda \big({\bf q}-f^2\, {\bf 1}_n\big)]
\end{align}
Written in a continuous form,
\begin{align}
&\sum_{\vec{\boldsymbol \sigma}}\exp{\Big[\alpha\beta^2\sum\limits_{a<b}r^{ab}\sum\limits_i\sigma_i^a\sigma_i^b+\frac{\beta}{2N}\sum\limits_a\sum\limits_{|i-j|<\frac{wN}{2}}\sigma_i^a\sigma_j^a\Big]}\nonumber\\
&=\int\prod_a\mathscr{D}\rho^a(x)\mathscr{D}\mu^a(x)\mathrm{d}\lambda^a\exp\Big[N\sum\limits_a\beta\lambda^a(\int\mathrm{d} x\,\rho^a(x)-f))\nonumber\\
&\ \ -\beta\int\mathrm{d} x\,\rho^a(x)\mu^a(x)+N\int\mathrm{d} x\, \log Z(\{\mu^a(x),r^{ab}\})\nonumber\\
&\ \ +\frac{\beta }{2}\int\mathrm{d} x\,\mathrm{d} y\,\rho^a(x)J_w(x-y)\rho^a(y)\Big]
\end{align}

where we have defined
\begin{equation}
\label{eq:Zaveraged}
 Z(\{\mu^a(x),r^{ab}\})\equiv\sum\limits_{\{\sigma^a\}}\exp[\alpha\beta^2\sum\limits_{a<b}\sigma^a\sigma^b r^{ab}+\beta\sum\limits_a\mu^a(x)\sigma^a]
\end{equation}

In the replica symmetric (RS) Ansatz, we assume 
\begin{equation}\forall\ a\neq b,\ \forall\ x,\ \left\{
\begin{array}{l}
r^{ab}=r\\
q^{ab}=q\\
\rho^a(x)=\rho(x)\\
\mu^a(x)=\mu(x)\\
\lambda^a=\lambda
\end{array}
\right.
\end{equation}

Under this assumption, we have
\begin{equation}
T_j=(n-1)(f-q)^j+(f-f^2+(n-1)(q-f^2))^j,
\end{equation}
and $Z(\{\mu^a(x),r^{ab}\})$ takes the expression 
\begin{equation}
 Z(\mu(x),r)=\int\mathrm{D}z\left[1+e^{\beta z\sqrt{\alpha r}+\beta\mu(x)-\frac{\alpha\beta^2r}{2}}\right]^n,
\end{equation}
so
\begin{align}
 \log Z(\mu(x),r) &= n\int\mathrm{D}z\log (1+e^{\beta z\sqrt{\alpha r}+\beta\mu(x)-\frac{\alpha\beta^2r}{2}})\nonumber\\
&+\frac{n^2}{2}\Big[\int\mathrm{D}z\log^2(1+e^{\beta z\sqrt{\alpha r}+\beta\mu(x)-\frac{\alpha\beta^2r}{2}})\nonumber\\
&-\big(\int\mathrm{D}z\log(1+e^{\beta z\sqrt{\alpha r}+\beta\mu(x)-\frac{\alpha\beta^2r}{2}})\big)^2\Big]+\mathcal{O}(n^3)
\end{align}

Under the RS hypothesis, the averaged partition function has the form:
\begin{equation}
\overline{Z^n}\sim\int\mathrm{d}q\ \mathrm{d}r\ \mathrm{d}\lambda\ \mathscr{D}\mu(x)\mathscr{D}\rho(x)e^{-nN\beta\hat{\mathscr{F}}[\mu(x),\rho(x),q,r,\lambda]}
\end{equation}
where

\begin{align}
\label{eq:exprG}
&\hat{\mathscr{F}}[\mu(x),\rho(x),q,r,\lambda,n]=-\frac1n\sum_j\frac{\beta^j}{2j}\Lambda_j T_j+(n-1)\frac{\alpha\beta}{2}rq\nonumber\\
&\ \ \ \ \ -\lambda\big(\int\mathrm{d} x\rho(x)-f\big)+\int\mathrm{d}x\rho(x)\mu(x)\nonumber\\
&\ \ \ \ \ -\frac 1{n\beta}\int\mathrm{d}x \log Z(\mu(x),r)-\frac{1}{2}\int\mathrm{d}x\mathrm{d}y\rho(x)J_w(x-y)\rho(y)
\end{align}

For large $N$, the saddle point approximation gives
\begin{equation}
 \overline{Z^n}\approx e^{-nN\beta\hat{\mathscr{F}}^*}
\end{equation}
where $\hat{\mathscr{F}}^*$ is an extremum of $\hat{\mathscr{F}}$ over all $\{\mu(x)\}$, $\{\rho(x)\}$, $q$, $r$, $\lambda$.\\

The replica trick then allows to compute the density of free energy $\mathscr{F}$ from the first term in $\hat{\mathscr{F}}$ as $n\rightarrow 0$: Letting 
\begin{equation}\psi(q,\beta)\equiv\sum_j\frac12\frac{\beta^j}{j}\Lambda_j[j(q-f)^2(f-q)^{j-1}+(f-q)^j],\end{equation}
 we obtain
\begin{align}
 \mathscr{F}&\equiv-\frac{1}{N\beta}\log\overline{Z}\nonumber\\
&=\frac{\alpha\beta}{2}r(f-q)-\frac{\alpha}{\beta}\psi(q,\beta)-\lambda\big(\int\mathrm{d} x\rho(x)-f\big)\nonumber\\
&+\int\mathrm{d}x\rho(x)\mu(x)-\frac{1}{2}\int\mathrm{d}x\mathrm{d}y\ \rho(x)J_w(x-y)\rho(y)\nonumber\\
&-\frac{1}{\beta}\int\mathrm{d}x\mathrm{D}z\log \big(1+e^{\beta\sqrt{\alpha r}z+\beta\mu(x)}\big)
\end{align}
where we have changed $\mu(x)\to\mu(x)-\frac{\alpha\beta r}{2}$.
At the saddle point, which is found by writing
\begin{align}
     \frac{\partial\mathscr{F}}{\partial q}=\frac{\partial\mathscr{F}}{\partial r}=\frac{\partial\mathscr{F}}{\partial \rho(x)}=\frac{\partial\mathscr{F}}{\partial \mu(x)}=\frac{\partial\mathscr{F}}{\partial \lambda}=0\ ,
\end{align}  
we have
\begin{align}
\label{eq:saddlepoint}
&r=-\frac{2}{\beta^2}\frac{\partial\psi}{\partial q},\nonumber\\
&q=\int\mathrm{d}x\int\mathrm{D}z[1+e^{-\beta z\sqrt{\alpha r}-\beta\mu(x)}]^{-2},\nonumber\\
&\rho(x)=\int\mathrm{D}z[1+e^{-\beta z\sqrt{\alpha r}-\beta\mu(x)}]^{-1},\nonumber\\
&\mu(x)=\int\mathrm{d}yJ_w(x-y)\rho(y)+\lambda,\nonumber\\
&\int\mathrm{d}x\rho(x)=f\ .
\end{align}
In $D=1$ dimension,
\begin{equation}
\Lambda^\text{1D}_j=2\sum\limits_{k\geq1}\left(\frac{\sin(\pi k w)}{\pi k}\right)^j\ .
\end{equation}
Defining $A_k\equiv\frac{\pi k}{\sin(\pi k w)}$, we have
\begin{equation}
 \psi^{\text{1D}}(q,\beta)=\sum\limits_{k\geq 1}\frac{\beta(q-f^2)}{A_k-\beta(f-q)}-\log\Big(1-\frac{\beta(f-q)}{A_k}\Big)
\end{equation}
and
\begin{equation}
 r^{\text{1D}}=2(q-f^2)\sum\limits_{k\geq 1}\big[A_k-\beta(f-q)\big]^{-2}
\end{equation}
In $D=2$ dimensions, defining
\begin{align}
 \phi(k_1,k_2)\equiv\frac{k_1k_2\pi^2}{\sin(k_1\pi \sqrt{w})\sin(k_2\pi \sqrt{w})}\ ,
\end{align}
we have
\begin{equation}
 \Lambda^\text{2D}_j=4\sum_{\underset{\neq(0,0)}{k_1,k_2}}\bigg(\frac{1}{\phi(k_1k_2)}\bigg)^j\ ,
 \end{equation}
so
\begin{align}
\psi^{\text{2D}}(q,\beta)=2\sum_{\underset{\neq(0,0)}{(k_1,k_2)}}&\bigg(\frac{\beta(q-f^2)}{\phi(k_1,k_2)-\beta(f-q)}\nonumber\\
&-\log\Big(1-\frac{\beta(f-q)}{\phi(k_1,k_2)}\Big)\bigg)\ ,
\end{align}
and
\begin{equation}
r^{\text{2D}}=4(q-f^2)\sum_{\underset{\neq(0,0)}{(k_1,k_2)}}\big(\phi(k_1,k_2)-\beta(f-q)\big)^{-2}\ .
\end{equation}

\section{Silent cells hypothesis - calculation of the free energy}\label{appact}

We now consider the hypothesis that only a fraction $c$ of cells are involved in each environment's representation (see main text). The partition function is now averaged over two types of disorder: the random permutation of the place field centers, as before, and the choice of the subset of cells participating in each map $\ell$, \emph{i.e.} the value of the random variables $\tau_i^{\ell}$:
\begin{align}\label{eq:znpartial}
 \overline{Z^n}=&\sum\limits_{\vec{\boldsymbol \sigma}}\exp[\beta\sum\limits_a\sum\limits_{i<j}J_{ij}^0\tau_i^0\tau_j^0\sigma_i^a\sigma_j^a]\nonumber\\
&\times\left\langle\exp[\beta\sum\limits_{\ell=1}^L\sum\limits_a\sum\limits_{i<j}J_{ij}^0\tau_{\pi(i)}^{\ell}\tau_{\pi(j)}^{\ell}\sigma_{\pi(i)}^a\sigma_{\pi(j)}^a]\right\rangle_{\boldsymbol \pi,\boldsymbol \tau}\ ,
\end{align}
where $\boldsymbol \tau$ denotes one realization of the $L\times N$ random variables $\tau_i^\ell$, and the $\tau_i^0$ are 1 if $i$ is a multiple of $1/c$ (in other terms, the place field centers for the reference environment are evenly spaced). The sum is now taken over configurations $\vec{\boldsymbol \sigma}$ satisfying two contraints:

\begin{equation}
 \left\{
\begin{array}{c}
        \frac{1}{N}\sum\limits_i\sigma_i^a=f\ \ \forall\ a \\
        \frac{1}{cN}\sum\limits_i \tau_i^{\ell}\sigma_i^a=f\ \ \forall\ a,\ell
       \end{array}
\right.\ .
\end{equation}

Using the function $\mathbb{1}(x)=1\ \text{if}\ x=0$ and 0 elsewhere, we write
\begin{align}
 \overline{Z^n}=C\sum\limits_{\text{all}\ \vec{\boldsymbol \sigma}}\mathbb{1}\Big(\frac{1}{N}\sum\limits_i\sigma_i^a-f\Big)e^{\beta\sum\limits_a\sum\limits_{i<j}J_{ij}^0\tau_i^0\tau_j^0\sigma_i^a\sigma_j^a}\chi(\vec{\boldsymbol\sigma})^L\ ,
\end{align}
where $C$ is a constant and in each environment $\ell$, 
\begin{align}
\chi(\vec{\boldsymbol\sigma})\equiv\left\langle\mathbb{1}\Big(\frac{1}{cN}\sum\limits_i \tau_i^{\ell}\sigma_i^a-f\Big)\cdot\xi(\vec{\boldsymbol\sigma},\boldsymbol c^{\ell})\right\rangle_{\boldsymbol \tau} , 
\end{align}
with
\begin{align}
\xi(\vec{\boldsymbol\sigma},\boldsymbol \tau^{\ell})\equiv\left\langle e^{-\beta\sum\limits_a\sum\limits_{i<j}J_{ij}^0\tau_{\pi(i)}^{\ell}\tau_{\pi(j)}^{\ell}(\sigma_{\pi(i)}^a-f)(\sigma_{\pi(j)}^a-f)}\right\rangle_{\boldsymbol\pi} . 
\end{align}
We can drop the $\ell$ index since we will average over $\boldsymbol \tau$.
The computation of $\xi(\vec{\boldsymbol\sigma},\boldsymbol \tau)$ follows exactly the same steps as before and, using the same notations as in Appendix \ref{appaverage}, we end up with
\begin{align}
\xi(\vec{\boldsymbol\sigma},\boldsymbol \tau)=\exp\Big[-\sum_{\lambda \ne 0} \text{Trace} \;\log[\text{\bf Id}_n - \beta \lambda \big({\bf \tilde{q}}-cf^2\, {\bf 1}_n\big)]\Big]\ ,
\end{align}
where $\bf \tilde{q}$ is now the $n\times n$ matrix of elements
\begin{align}
 \tilde{ q}_{ab}\equiv\frac1N\sum\limits_i \tau_i\sigma_i^a\sigma_i^b\ .
\end{align}
We can now calculate $\chi(\vec{\boldsymbol\sigma})$: introducing parameters $R_{ab}$ conjugated to $\tilde{q}_{ab}$  and Lagrange multipliers $\lambda_a$ to enforce the constraint on $\vec{\boldsymbol \sigma}$, and letting
\begin{align}
 \Theta({\bf \tilde{q}})\equiv-\sum_{\lambda \ne 0} \text{Trace} \;\log[\text{\bf Id}_n - \beta \lambda \big({\bf \tilde{q}}-cf^2\, {\bf 1}_n\big)]\ ,
\end{align}
we can write
\begin{align}
 \chi(\vec{\boldsymbol\sigma})=\int\limits_{i\mathbb{R}}&\frac{\mathrm{d}\lambda_a}{2\pi\sqrt{N}}\frac{\mathrm{d}R_{ab}}{2\pi\sqrt{N}}\frac{\mathrm{d}\tilde{q}_{ab}}{2\pi}\cdot e^{Ncf\sum\limits_a\frac{\lambda_a}{\sqrt{N}}+\sqrt{N}\sum\limits_{a<b}R_{ab}\tilde{q}_{ab}+\Theta({\bf \tilde{q}})}\nonumber\\
&\times\prod\limits_i\left\langle \exp\big[-\tau_i(\sum\limits_a\frac{\lambda_a}{\sqrt{N}}\sigma_i^a+\frac{R_{ab}}{\sqrt{N}}\sigma_i^a\sigma_i^b)\big]\right\rangle_{\tau_i}.
\end{align}
Thus the $i$ are decoupled and we can perform the average over the $\tau_i$:
the averaged term at each $i$ is
\begin{align}
1-&c+c\exp\big[-\sum\limits_a\frac{\lambda_a}{\sqrt{N}}\sigma_i^a+\frac{R_{ab}}{\sqrt{N}}\sigma_i^a\sigma_i^b\big]\nonumber\\
&\sim\exp\big[-\sum\limits_a\frac{\lambda_a}{\sqrt{N}}\sigma_i^a+\frac{R_{ab}}{\sqrt{N}}\sigma_i^a\sigma_i^b\big]\nonumber\\
&\ \ \ \times\exp\big[\frac {c(1-c)}{2}\big(\sum\limits_a\frac{\lambda_a}{\sqrt{N}}\sigma_i^a+\frac{R_{ab}}{\sqrt{N}}\sigma_i^a\sigma_i^b\big)^2\big]
\end{align}
in the large $N$ limit. Introducing the notations 
\begin{align}
&T_{abc}\equiv\frac1N\sum\limits_{i}\sigma_i^a\sigma_i^b\sigma_i^c\nonumber\ ,\\ &S_{abcd}\equiv\frac1N\sum\limits_{i}\sigma_i^a\sigma_i^b\sigma_i^c\sigma_i^d\ ,
\end{align}
we have
\begin{align}
 &\chi(\vec{\boldsymbol\sigma})=\int\limits_{i\mathbb{R}} \exp\big[\sum\limits_{a<b}\Theta({\bf \tilde{q}})+\sqrt{N}R_{ab}(\tilde{q}_{ab}-cq_{ab})\big]\nonumber\\
&\times \exp\big[\frac{c(1-c)}{2}\big(\sum\limits_{a,b,c,d}\lambda_a\lambda_bq_{ab}+\lambda_aR_{bc}T_{abc}
+R_{ab}R_{cd}S_{abcd}\big)\big].
\end{align}
After Gaussian integration on the $\lambda_a$ and $R_{ab}$, we end up with (up to a multiplicative constant)
\begin{align}
 &\chi(\vec{\boldsymbol\sigma})\sim\int\limits_{i\mathbb{R}} \frac{\mathrm{d}\tilde{q}_{ab}}{2\pi}\exp\big[\sum\limits_{a<b}\Theta({\bf \tilde{q}})\big]\\
&\times \exp\big[\frac{N}{2}\sum\limits_{a,b,c,d}[A^{-1}]_{abcd}(\tilde{q}_{ab}-cq_{ab})(\tilde{q}_{cd}-cq_{cd})\big] ,\nonumber
\end{align}
where
\begin{align}
 A_{abcd}\equiv c(1-c)(S_{abcd} +\frac14[Q^{-1}]_{ab}T_{acd}T_{bcd})\ .
\end{align}

Hence, in the large $N$ limit, the integral is dominated by $\tilde{q}_{ab}\sim cq_{ab}$.\\
Then we write the partition function  and do the replica symmetric Ansatz as in Appendix \ref{appreplica}. The difference is that now $\mu(x)$ and $\rho(x)$ describe the activity of cells involved in the reference environment (a dilution factor $c$ appears), the $(1-c)N$ other cells having a uniform activity. We thus derive the expression of the energy functional (\ref{eq:Fc}) and the saddle point equations (\ref{eq:saddlepoint2}).

\section{Stability of the replica symmetric solution}
\label{appstab}
The extremization of the free energy functional under the fixed-activity constraint and under the replica symmetric assumption leads to three solutions corresponding to three different phases.
 We want to study the stability of those solutions in the $(\alpha, T)$ space. We will  limit ourselves to the one-dimensional case.
For this purpose we do a small perturbation of the solution and observe the behaviour of the free energy functional:\\ If
\begin{equation}
\left\{
\begin{array}{ccc}
\rho^a(x)&\rightarrow&\rho^a(x)+\delta\rho^a(x)\\
 \mu^a(x)&\rightarrow&\mu^a(x)+\delta\mu^a(x)\\
r^{ab}&\rightarrow&r^{ab}+\delta r^{ab}\\
q^{ab}&\rightarrow&q^{ab}+\delta q^{ab}
\end{array}
\right. 
\end{equation}
 then $\mathscr{F}\rightarrow\mathscr{F}+\underbrace{\delta\mathscr{F}}_{=0}+\frac{1}{2}\delta^2\mathscr{F}$\ ,\\
 where 
\begin{equation}
 \delta^2\mathscr{F}=\int\mathrm{d}x\mathrm{d}y\left[\begin{array}{c}\delta\rho^a(x)\\\delta\mu^a(x)\\\delta r^{ab}\\\delta q^{ab}\end{array}\right]^\dagger\cdot M(x,y)\cdot\left[\begin{array}{c}\delta\rho^c(y)\\\delta\mu^c(y)\\\delta r^{cd}\\\delta q^{cd}\end{array}\right]\ .
\end{equation}
We thus have to study the hessian matrix $M(x,y)$ that writes, in the $\{\delta\rho^a(x),\delta\mu^a(x),\delta r^{ab},\delta q^{ab}\}$ basis:

\begin{equation}
M=\left[
\begin{array}{c|c|c|c}

&&&\\
\ \frac{\partial^2\mathscr{F}}{\partial \rho^{a}(x)\partial \rho^{c}(y)}\ &\frac{\partial^2\mathscr{F}}{\partial \rho^{a}(x)\partial \mu^{c}(y)}\ &0&0\\ 
&&&\\ \hline
&&&\\
\frac{\partial^2\mathscr{F}}{\partial \mu^{a}(x)\partial \rho^{c}(y)}&\frac{\partial^2\mathscr{F}}{\partial \mu^{a}(x)\partial \mu^{c}(y)}&\frac{\partial^2\mathscr{F}}{\partial \mu^{a}(x)\partial r^{cd}}&0\\ 
&&&\\ \hline
&&&\\
0&\frac{\partial^2\mathscr{F}}{\partial r^{ab}\partial \mu^{c}(y)}&\frac{\partial^2\mathscr{F}}{\partial r^{ab}\partial r^{cd}}&\frac{\partial^2\mathscr{F}}{\partial r^{ab}\partial q^{cd}}\\ 
&&&\\ \hline
&&&\\
0&0&\frac{\partial^2\mathscr{F}}{\partial q^{ab}\partial r^{cd}}&\frac{\partial^2\mathscr{F}}{\partial q^{ab}\partial q^{cd}}\\
&&&
\end{array}
\right]
\end{equation}

where the expressions of the elements of the different blocks are detailed hereafter: using the notations
\begin{align}
&t(x)\equiv\overline{\langle\sigma\rangle^3_{(x)}}=\int\mathrm{D}z[1+e^{-\beta z\sqrt{\alpha r}-\beta\mu(x)}]^{-3}\ ,\nonumber\\
&s(x)\equiv\overline{\langle\sigma\rangle^4_{(x)}}=\int\mathrm{D}z[1+e^{-\beta z\sqrt{\alpha r}-\beta\mu(x)}]^{-4}\ ,\nonumber\\
&t\equiv\int\mathrm{d}x\ t(x)\ ;\ s\equiv\int\mathrm{d}x\ s(x)\ ;\ q_2\equiv\int\mathrm{d}x\ q^2(x)\ ,
\end{align}
we have

\begin{flalign}
 \frac{\partial^2\mathscr{F}}{\partial \rho^{a}(x)\partial \rho^{c}(y)}=-J_w(x-y)\cdot\delta^{ab}\ ,
\end{flalign}

\begin{flalign}
 \frac{\partial^2\mathscr{F}}{\partial \rho^{a}(x)\partial \mu^{c}(y)}=\delta(x-y)\cdot\delta^{ab},
\end{flalign}
\begin{equation}
\frac{\partial^2\mathscr{F}}{\partial \mu^{a}(x)\partial \mu^{c}(y)}=
\left\{
\begin{array}{ll}
\delta(x-y)\beta(\rho^2(x)-\rho(x))&\text{if}\ a=b\\
\delta(x-y)\beta(\rho^2(x)-q(x))&\text{otherwise}
 \end{array}
\right. ,
\end{equation}
\begin{equation}
\frac{\partial^2\mathscr{F}}{\partial \mu^{a}(x)\partial r^{cd}}=
\left\{
\begin{array}{ll}
\alpha\beta^2(q(x)\rho(x)-t(x))&\text{if}\ a\neq c\neq d\\
\alpha\beta^2(q(x)\rho(x)-q(x))&\text{otherwise}
 \end{array}
\right. ,
\end{equation}
\begin{equation}
\frac{\partial^2\mathscr{F}}{\partial r^{ab}\partial r^{cd}}=
\left\{
\begin{array}{ll}
\alpha^2\beta^3(\int q^2 -q)&\text{if}\ a=c\ \text{and}\ b=d\\
\alpha^2\beta^3(\int q^2 -s)&\text{if}\ a\neq b\neq c\neq d\\
\alpha^2\beta^3(\int q^2 -t)&\text{otherwise}
 \end{array}
\right. ,
\end{equation}
and, letting 
\begin{align*}
&B_k\equiv\frac{k\pi}{\sin(k\pi w)}-\beta(f-q)\ ,\\
&C_1\equiv\sum\limits_{k\ge1}\frac{\beta}{B_k^2}\ ,\\
&C_2\equiv\sum\limits_{k\ge1}\frac{\beta^2(q-f^2)}{B_k^3}\ ,\\
&C_3\equiv\sum\limits_{k\ge1}\frac{\beta^3(q-f^2)^2}{B_k^4}\ ,
\end{align*}
\begin{equation}
\frac{\partial^2\mathscr{F}}{\partial q^{ab}\partial q^{cb}}= \left\{
\begin{array}{ll}
-2\alpha(C_1+2C_2+2C_3) & \text{if}\ a=c\ \text{and}\ b=d\\
-4\alpha C_3 & \text{if}\ a\neq b \neq c \neq d\\
-2\alpha(C_2+2C_3) & \text{otherwise}
\end{array}
\right.\ .
\end{equation}

The eigenvector equation writes
\begin{equation}
\label{eq:valeurspropres}
 M\cdot\vec{v}=\lambda\cdot\vec{v}
\end{equation}
where $\vec{v}$ is the vector of fluctuations around the saddle point:
\begin{equation}
 \vec{v}(x)=\left[
\begin{array}{c}
                   \delta\rho^a(x)\\
\vdots\\
\delta\mu^a(x)\\
\vdots\\
\delta r^{ab}\\
\vdots\\
\delta q^{ab}\\
\vdots\\
                  \end{array}
\right]\ .
\end{equation}

Following the same strategy as \cite{deAlmeidaThouless78} to exploit the symmetry of the matrix elements under permutation of the indices, we look for orthogonal set of eigenvectors with the particular forms below

\begin{equation}
\vec{v_1}(x) = \left\{\begin{array}{cl}
\delta\rho^a(x)=\delta{\rho}(x) & \forall a\\
\delta\mu^a(x)=\delta{\mu}(x) & \forall  a\\
\delta r^{ab}=\delta{r} & \forall a,b\\
\delta q^{ab}=\delta{q} & \forall a,b
\end{array}
\right.\ , 
\end{equation}

\begin{equation}
\vec{v_2}(x) = \left\{
\begin{array}{cccl}
\delta\rho^a(x)&=&\delta\hat{\rho}(x) & \mathrm{if}\ a=\theta\ \\
&=&\delta\check{\rho}(x) & \mathrm{otherwise}\\
\delta\mu^a(x)&=&\delta\hat{\mu}(x) & \mathrm{if}\ a=\theta\ \\
&=&\delta\check{\mu}(x) & \mathrm{otherwise}\\
\delta r^{ab}&=&\delta\hat{r} & \mathrm{if}\ a\ \mathrm{or}\ b=\theta\\
&=&\delta\check{r} & \mathrm{if}\ a\ \mathrm{and}\ b\neq\theta\\\
\delta q^{ab}&=&\delta\hat{q} & \mathrm{if}\ a\ \mathrm{or}\ b=\theta\\
&=&\delta\check{q} & \mathrm{if}\ a\ \mathrm{and}\ b\neq\theta
\end{array}
\right.\ ,
\end{equation}

\begin{equation}
\vec{v_3}(x) = \left\{
\begin{array}{cccl}
\delta\rho^a(x)&=&\delta\tilde{\rho}(x) & \mathrm{if}\ a=\theta\ \mathrm{or}\ \theta'\\
&=&\delta\rho^*(x) & \mathrm{otherwise}\\
\delta\mu^a(x)&=&\delta\tilde{\mu}(x) & \mathrm{if}\ a=\theta\ \mathrm{or}\ \theta'\\
&=&\delta\mu^*(x) & \mathrm{otherwise}\\
\delta r^{ab}&=&\delta\tilde{r} & \mathrm{if}\ a=\theta\ \mathrm{and}\ b=\theta'\\
&=&\delta\tilde{\tilde{r}} & \mathrm{if}\ a\ \mathrm{or}\ b=\theta\ \mathrm{or}\ \theta'\\
&=&\delta r^* & \mathrm{if}\ a\ \mathrm{and}\ b\neq\theta, \theta'\\
\delta q^{ab}&=&\delta\tilde{q} & \mathrm{if}\ a=\theta\ \mathrm{and}\ b=\theta'\\
&=&\delta\tilde{\tilde{q}} & \mathrm{if}\ a\ \mathrm{or}\ b=\theta\ \mathrm{or}\ \theta'\\
&=&\delta q^* & \mathrm{if}\ a\ \mathrm{and}\ b\neq\theta, \theta'
\end{array}
\right.\ ,
\end{equation}
where $\theta$ and $\theta'$ are two fixed replica indices. $\vec{v_1}(x)$ and $\vec{v_2}(x)$ are called longitudinal modes; $\vec{v_3}(x)$ are called transverse or 'replicon' modes. 

Imposing the orthogonality conditions 
\begin{equation}
\vec{v_1}(x)\cdot\vec{v_2}(x)=\vec{v_1}(x)\cdot\vec{v_3}(x)=\vec{v_2}(x)\cdot\vec{v_3}(x)=0\ ,
\end{equation}

and taking the $n\rightarrow 0$ limit in equations (\ref{eq:valeurspropres}), we end up with two systems of equations: on the longitudinal modes the eigenvalues equation leads to

\begin{equation}
\label{eq:valeurspropressimplifiee}
 \left\{
\begin{array}{l}
-\int\mathrm{d}y\,J_w(x-y)\delta\rho(y)+\delta\mu(x)=\lambda\ \delta\rho(x)\\
 \\

\delta\rho(x)+\beta(q-\rho)(x)\delta\mu(x)+\alpha\beta^2(q-t)(x)\delta r =\lambda\ \delta\mu(x)\\
 \\

2\alpha\beta^2\int\mathrm(t-q)\delta\mu+\alpha^2\beta^3(-q+4t-3s)\delta r+\alpha\beta\delta q=\lambda\ \delta r\\
 \\
\alpha\beta\delta r-2\alpha(C_1-2C_2)\delta q=\lambda\ \delta q
\end{array}
\right.\ ,
\end{equation}
and on the transverse modes it gives 
\begin{equation}
\label{eq:replicons}
\left\{
\begin{array}{l}
\alpha^2\beta^3[-q+2t-s]\delta r^*+\alpha\beta\delta q^*=\lambda\ \delta r^*\\
\alpha\beta\delta r^*-2\alpha C_1\delta q^*=\lambda\ \delta q^*
\end{array}
\right.\ .
\end{equation}

For each of the three phases determined above,  the stability regions in the $(\alpha,T)$ domain are delimited by lines where one of the eigenvalues changes signs.\\
Note that the matrix of system (\ref{eq:valeurspropressimplifiee}) is not symmetric while the hessian matrix $\delta^2\mathscr{F}$ is: a $-\frac 12$ factor appears when taking the $n\rightarrow 0$ limit since there are $\frac{n(n-1)}{2}$ two-replica-indice components. This multiplicative factor does not change the points where a given eigenvalue changes signs. Nevertheless, it has the effect of giving nonreal eigenvalues. To avoid this, we shall restore the symmetry by multiplying the appropriate lines by $-\frac 12$.

\subsection{Paramagnetic phase stability region}\label{appstabpara}
Taking $\rho(x)=f$, $q(x)=f^2$, $t(x)=f^3$ and $s(x)=f^4$ for all $x$ in (\ref{eq:valeurspropressimplifiee}) leads to a very simple system, invariant under translation in the $x$ space.\\
We can solve it by Fourier transform: if
\begin{equation}
 \begin{array}{ll}
  \delta{\rho}(x)&=\sum\limits_{k>0}e^{2i\pi kx}\delta{\rho}(k)\\
\delta{\mu}(x)&=\sum\limits_{k\geq0}e^{2i\pi kx}\delta{\mu}(k)\\
 \end{array}
\end{equation}
($\delta\rho(0)=0$ is imposed by the fixed activity constraint),
then the system (\ref{eq:valeurspropressimplifiee}) decomposed on each Fourier mode  gives:
\begin{enumerate}

\item \underline{$k>0$ components of the longitudinal matrix:} these modes appear in the $(\delta\rho(x),\delta\mu(x))$ region and solve a system with determinant
\begin{equation}
 \left|
\begin{array}{cc}
 -\frac{\text{sin}(\pi kw)}{\pi k}&1\\
1&\beta(f^2-f)
\end{array}
\right|
\end{equation}
that vanishes for $\beta(k)=\frac{\pi k}{\text{sin}(\pi k w)(f-f^2)}$ which is minimal for $k=1$. For $f=0.1$ and $w=0.05$, $T_1\approx0.0045$.

 \item \underline{$k=0$ component of the longitudinal matrix:} we get a system with determinant
\begin{equation}
 \left|
\begin{array}{ccc}
 f^2-f&\beta(f^2-f^3)&0\\
2\alpha\beta(f^3-f^2)&\alpha\beta^2(-f^2+4f^3-3f^4)&1\\
0&1&-2\frac{C_1}{\beta}\\
\end{array}
\right|
\end{equation}
These modes appear for $(\alpha,T_0(\alpha))$ at which this determinant vanishes, \emph{i.e.} 
\begin{equation}\label{eqlongipara}
 \sum\limits_{k\geq1}\left[\frac{{T}_0(\alpha)\,k \pi}{f(1-f)\sin(k\pi w)}-1\right]^{-2}=\frac{1}{2\alpha} \ .
\end{equation}

\item \underline{Replicon modes:} these modes solve a system with determinant
\begin{equation}
 \left|
\begin{array}{cc}
\alpha\beta^2(-f^2+2f^3-f^4)&1\\
1&-2\frac{C_1}{\beta}
\end{array}
\right|
\end{equation}
This defines the same stability line (\ref{eqlongipara}) as found above.
\end{enumerate}
To sum up, the paramagnetic phase is stable at high temperatures; when $T$ decreases at fixed $\alpha$, it becomes instable at $T_{\text{PM}}(\alpha)=\text{max}\left\{T_0(\alpha),T_1\right\}$ as depicted in Fig.~\ref{fig:stabpara}.

\subsection{Glassy phase stability region}\label{appstabglass}
We find a uniform solution to the saddle point equations (\ref{eq:saddlepoint1}) with $q>f^2$ only for $T<T_{\text{PM}}(\alpha)$: the region of existence of the glassy phase hence corresponds to the region where the PM solution is unstable. In this region we find that the RS solution is always stable against longitudinal modes (\ref{eq:valeurspropressimplifiee}) and always unstable against transverse modes (\ref{eq:replicons}). The replica symmetric Ansatz is therefore not correct in the case of the glassy phase. The correct expression could be found by looking for replica-symmetry broken solutions. Since we are chiefly interested by the clump phase, we skipped this computation and did not investigate further the glassy phase.

\subsection{Clump phase stability region}\label{appstabclump}
\begin{itemize}
\item \underline{longitudinal modes:} 
Due to the $x$ dependency in this phase, we must use a numerical approach in a discretized space to study the eigenvalues of the longitudinal matrix. Since computation time for the matrix diagonalization limits dramatically the number of points in the discretization, we chose to study the longitudinal stability with a different method. Scanning the $(\alpha,T)$ plane, $\rho(x)$ is computed by  solving iteratively the saddle-point equations (\ref{eq:saddlepoint1}) starting from a initial clump;  the line of stability corresponds to the points where the clump collapses, \emph{i.e.} the iteration converges to a uniform activity $\rho(x)=f\ \forall x$. The result is shown in Fig.~\ref{fig:stabclump} in the main text.
\item \underline{replicon modes:} 
For all $\alpha, T$, we compute numerically $q$, $t$, $s$ by solving iteratively the saddle-point equations as before, allowing to calculate the determinant of system (\ref{eq:replicons}). We looked for the line where this determinant vanishes. We found that replica symmetry breaking is limited to  a small region at the low $T$/ high $\alpha$ edge of the region of longitudinal stability, see Fig.~\ref{fig:stabclump}.
\end{itemize}

\bibliography{refplacecells}
\bibliographystyle{apsrev4-1}

\end{document}